\documentclass[11pt,a4paper]{article}

\usepackage{amsmath}
\usepackage{amsfonts}
\usepackage{amsthm}
\usepackage{graphicx}

\def\Z{\mathbb Z}
\def\C{\mathbb C}
\def\R{\mathbb R}

\def\dom{\mathrm {dom}\,}
\def\LL{\mathrm L}
\def\CC{\mathrm C}

\def\KK{\mathrm K}

\newcommand{\la}{\langle }
\newcommand{\ra}{\rangle }

\newtheorem{teor2}{Theorem}

\newtheorem{prop2}{Proposition}
\newtheorem{lema2}{Lemma}
\newtheorem{obs2}{Remark}

\theoremstyle{definition}
\newtheorem{exem2}{Example}
\newtheorem{defi2}{Definition}

\newcommand{\img}{{\mathrm{rng}~}}

\newcommand{\Id}{{\mathbf 1}}

\newcommand{\hil}{\mathcal H}

\newcommand{\ddd}{:=}

\begin{document}

 \title{Scattering and self-adjoint extensions of the Aharonov-Bohm
hamiltonian}
\author{C\'{e}sar R. de Oliveira \\
\vspace{-0.6cm}
\small
\it Departamento de Matem\'{a}tica -- UFSCar, \small \it S\~{a}o Carlos,
SP, 13560-970 Brazil\\ \\ \\
Marciano Pereira\thanks{Corresponding author. Email: marciano@uepg.br} \\
\vspace{-0.6cm}
\small
\it Departamento de Matem\'{a}tica e Estat{\'\i}stica -- UEPG,  \\ \\
\it \small Ponta Grossa, PR, 84030-000 Brazil\\ \\}
\date{\today}

\maketitle

\begin{abstract} We consider the hamiltonian operator associated with
planar sections of infinitely long cylindrical solenoids and with
a homogeneous magnetic field in their interior. First, in the Sobolev space
$\hil^2$, we characterize all generalized boundary conditions on the
solenoid border compatible with quantum mechanics, i.e., the boundary
conditions so that the corresponding hamiltonian operators are
self-adjoint. Then we study and compare the scattering of the most usual
boundary conditions, that is, Dirichlet, Neumann and Robin.
\end{abstract}

\vspace{1cm}
 \vspace{1cm}

\tableofcontents

\section{Introduction}
Although the Aharonov-Bohm (AB) effect   is a fundamental question in
quantum physics, and despite the original work on the AB effect has been
published 50 years ago \cite{AB59,PT}, it is still a very active area of
research with many open mathematical questions. Here we address some of
these questions and, our first aim is to try to characterize, in the
two-dimensional space,
all boundary conditions on the (cylindrical) solenoid border $\mathcal S$
that are compatible with quantum mechanics, and whose domains are
subspaces of the natural Sobolev space $\hil^2(\mathcal S')$, where
$\mathcal S'$ is the exterior region of the solenoid.

 Our (standard) cylindrical solenoid $\mathcal S$ has radius $a>0$, is
infinitely long and centered at the origin (with axis coinciding with the
$z$ direction), it carries a stationary electric current so that there is
a homogeneous magnetic field $\mathbf B=(0,0,B)$ confined to the solenoid
interior ${\mathcal S}^\circ$, and vanishing in the exterior region
$\mathcal S'$. A spinless charged  particle of mass $m=1/2$ lives in
$\mathcal S'$ and has no contact with the magnetic field $\mathbf B$. If
$\mathbf A$ is a vector potential that generates such magnetic field,
that is,  $\mathbf B= \nabla\times\mathbf A$, the initial (quantum) AB
hamiltonian operator for such charged particle  is given  by (with
$\hbar=1$, $c$ and $q$ stand for the speed of light and particle electric
charge, respectively)
\begin{equation}\label{initialABh} H = \left(\mathbf p - \frac qc \mathbf
A\right)^2,\quad
\mathbf p = -i\nabla,\qquad \dom H=\CC^\infty_0(\mathcal S').
\end{equation} Ahead, we will fix a specific choice of the vector potential.

 This operator $H$ is not self-adjoint and so does not
correspond to a
physical observable; the possible self-adjoint extensions characterize all
possible physical interaction of the particle with the solenoid border
(sometimes obtained through non-trivial limit procedures). It is important
to note that the elements $\psi$ of  the domain  $\CC^\infty_0(\mathcal
S')$ do not touch $\mathcal S$ (in the sense that $\psi=0$ in a
neighborhood of the solenoid) and $H\psi$ has a very simple action.  Such
collection of ``exotic'' extensions might, for instance, include the
description of limit situations where singular perturbations (that could
also depend on time) are slowly turned off.

It is usually assumed that the domain of the ``physical'' self-adjoint
extensions is Dirichlet (i.e., $\psi=0$ on $\mathcal S$), and some
theoretical arguments have appeared to justify such choice (see
\cite{deOP} and references therein). However, here we take an open-minded
position and ask about other possibilities of boundary conditions. In a
general sense, these other conditions correspond to the requirement of
vanishing of the probability current at the solenoid border, and they may
model different sorts of interactions between the particle and the
solenoid. In Section~\ref{SectionSAE} we characterize all of such boundary
conditions whose domain of the corresponding self-adjoint hamiltonians are
composed of functions with square integrable first and second derivatives
(these are rather natural technical conditions in quantum mechanics). The
operator $H$ above has deficiency indices equal to infinity, and its
self-adjoint
extensions should be compared with the case of solenoid of zero radius
discussed, for instance,  in \cite{AT,DaSt}, where the deficiency indices
are equal to~2.

In Section~\ref{SectionScatter} the two-dimensional scattering will be
discussed in this context, but restricted to the traditional boundary
conditions: Dirichlet, Neumann and Robin. In fact, the case of Dirichlet
was investigated in \cite{Ruij} and part of our results are based on the
techniques discussed therein. The scattering in case of solenoids with
zero radius is discussed in \cite{AB59,AT,DaSt,Hagen,PanRich}.

For solenoids of radius greater than zero, and the above mentioned
traditional extensions, we will show that the wave
operators exist and are complete; we  also find expressions for the
scattering operators and their asymptotic behaviours for low and high
energies. Finally, we will find explicitly the respective differential
scattering cross sections (an important ingredient in experiments) and
some figures will compare their values. From some point of view, the
discrepancy among such figures
could, in principle, be useful for an experimental selection of the
boundary condition
occurring in each situation; in principle it is not obvious which boundary
conditions are naturally realized in laboratories, and we have found that
given two of such self-adjoint extensions, it is always possible to find a
range of energy so that the corresponding scattering cross sections can be
distinguished.

\section{Self-adjoint extensions}\label{SectionSAE}In this section we find
and characterize an important class of  self-adjoint extensions of the
initial hamiltonian~$H$ (see equation \eqref{initialABh}), with vector
potential  ${\bf A}$  given, in polar
coordinates $(r,\theta)$, by ${\bf A} = (A_r,A_\theta)$, $A_r\equiv 0$ and
$A_\theta = \displaystyle\frac{\Phi}{2\pi r}$, $r \geq a$, and
$\Phi$ is the total magnetic flux through the solenoid. As will be
discussed in Section~\ref{SectionScatter}, this hermitian operator has
deficiency indices \cite{ISTQD} $n_+(H)=n_-(H)=+\infty$, and so it has
infinitely many
self-adjoint extensions.

The first physical and mathematical point to be addressed is to find some
self-adjoint operators that may potentially describe the Aharonov-Bohm
hamiltonian of a charged particle moving in the exterior region $\mathcal
S'$. This particle can not penetrate the solenoid  but interacts with
$\mathcal S$, and the boundary conditions that give rise to self-adjoint
realizations are the possible conditions, from the point of view of
quantum mechanics, that may describe such interaction with different types
of interface materials and limit procedures.

We note that in order to classify all such extensions it is necessary to
make use of some Sobolev spaces $\hil^s(\mathcal S)$ with
$s<0$ \cite{Adams}. Furthermore, another difficulty is that the domain of
the adjoint
operator
\begin{equation}
\dom H^*  =  \left\{\psi\in {\textrm L}^2(\mathcal S') : H\psi\in{\textrm
L}^2(\mathcal S')\right\}
\end{equation}
is not contained in the space $\hil^2(\mathcal S')$ \cite{G68, G06,
G08}, which has proved to be natural in quantum problems. Below we shall
restrict our arguments to the extensions whose domains are contained in
$\hil^2(\mathcal S')$, which will permit us to use boundary triples to
find these extensions in a quite simpler way (see Remark~\ref{remaHneg}).

Due to the symmetry of the problem, we shall consider a planar cross
section; another fact is that the solenoid border in  $\R^3$ is not a
compact set, and so it is not clear how to define the trace operators in
the spatial case (and we want to avoid this technical point).

The reader interested only in the final results may go straightly to
Theorem~\ref{mainTeor} and the examples that follow this theorem.

\subsection{Boundary triples}
Our way to find self-adjoint extensions of $H$ is via boundary triples
$({\bf h},\rho_1,\rho_2)$, as described in  \cite{ISTQD}, Chapter~7, so we
present a brief account of this technique.

\begin{defi2}
Let $T$ be a hermitian operator in a Hilbert space $\hil$. The {\em
boundary form} of $T$ is the sesquilinear mapping $\Gamma =
\Gamma_{T^\ast} : \dom T^\ast \times \dom T^\ast \to \C$ given by
\begin{equation}
\Gamma(\xi,\eta)\ddd \left\langle T^\ast \xi, \eta \right\rangle -
\left\langle \xi, T^\ast \eta \right\rangle,\quad \xi,\eta\in\dom T^\ast.
\end{equation}
\end{defi2}
\begin{prop2}
$\Gamma(\xi, \eta) = 0$, for all $\xi,\eta\in\dom T^\ast$, if, and only
if, $T^\ast$ is self-adjoint, that is, if, and only if, $T$ is essentially
self-adjoint.
\end{prop2}

Boundary forms can be used to determine self-adjoint extensions of $T$ by
noting that such extensions are restrictions of $T^\ast$ to certain
domains $\mathcal D$ such that $\Gamma(\xi, \eta)= 0$, for all
$\xi,\eta\in \mathcal D$.
By von Neumann theory \cite{ISTQD}, each self-adjoint extension of $T$ is
in a one-to-one correspondence with unitary operators  $\hat U:\KK_-(T)\to
\KK_+(T)$ between the deficiency subspaces $\KK_\pm$ of $T$; denote by
$T^{\hat U}$ the corresponding self-adjoint extension whose domains is
($\overline T$ denotes the closure of the operator~$T$)
\begin{equation}
\dom T^{\hat U} = \{ \eta = \zeta + \eta_- - \hat U\eta_- : \zeta\in\dom
\overline T, \eta_-\in\KK_-(T) \}.
\end{equation}
Note that the boundary form $\Gamma$ restricted to $\dom T^{\hat U}$
vanishes.

Now we recall the concept of boundary triples.

\begin{defi2}
Let $T$ be a hermitian operator with deficiency indices $n_-(T)=n_+(T)$. A
{\em boundary triple} $(\mathbf h, \rho_1, \rho_2)$ for $T$ is composed of
a Hilbert space $\mathbf h$ and two linear mappings $\rho_1, \rho_2 : \dom
T^\ast\to \mathbf h$ with dense images and so that
\begin{equation}
b\,\Gamma_{T^\ast}(\xi, \eta) = \left\langle \rho_1(\xi), \rho_1(\eta)
\right\rangle - \left\langle \rho_2(\xi), \rho_2(\eta) \right\rangle,
\quad \forall\,\xi, \eta\in \dom T^\ast,
\end{equation}
for some constant $0\neq b \in \C$. $\left\langle\cdot,\cdot\right\rangle$
denotes the inner product in  $\mathbf h$ and $\dim \mathbf h = n_+(T)$.
\end{defi2}

Again, self-adjoint extensions of $T$ are restrictions of $T^\ast$ to
certain domains $\mathcal D$ so that $\Gamma(\xi, \eta) = 0$, for all
$\xi, \eta\in\mathcal D$, and given a boundary triple for $T$, such
domains $\mathcal D$ are related to unitary operators $U:\mathbf h \to
\mathbf h$ so that $U\rho_1(\xi) = \rho_2(\xi)$ and
\begin{equation}
\left\langle \rho_1(\xi), \rho_1(\eta) \right\rangle = \left\langle
\rho_2(\xi), \rho_2(\eta) \right\rangle = \left\langle U\rho_1(\xi),
U\rho_1(\eta) \right\rangle,\quad \forall\,\xi,\eta\in\mathcal D.
\end{equation}

The main results we need here are summarized in the following theorem.

\begin{teor2}\label{extviatriplas}
Let $T$ be a hermitian operator with equal deficiency indices. If
$(\mathbf h, \rho_1, \rho_2)$ is a boundary triple for $T$, then the
self-adjoint extensions of $T$ are given by
\begin{equation}
\begin{split}
\dom T^U & = \left\{ \xi\in\dom T^\ast : \rho_2(\xi) = U\rho_1(\xi)
\right\}, \\
T^U\xi & = T^\ast \xi,\qquad\xi\in\dom T^U,
\end{split}
\end{equation}
for each unitary operator $U:\mathbf h\to\mathbf h$.
\end{teor2}

\subsection{Boundary triples for the AB operator}
Some self-adjoint extensions of the initial AB
operator $H$ will be found. It will combine the cylindrical symmetry with
the topological
property of multiply connectedness, that is,  the plane with a circular
hole, without mentioning the important ingredient of a magnetic potential
$\mathbf A$ with $\mathrm{div\,} \mathbf A = 0$ in $\mathcal S'$.   The
method can be
adapted to other regions with boundaries so that the trace construction
applies (e.g., smooth and compact boundaries).

Although a $\psi(r,\theta)\in\hil^1({\mathcal S'})$ is not necessarily
continuous, it is possible to give a meaning to the
restriction
$\psi(a,\theta)=\psi|_{\mathcal S}(\theta)\in {\LL}^2(\mathcal S)$ via the
so-called {\em trace} (more properly, it
should be called {\em Sobolev trace}) of
$\psi$; see ahead. It turns out that there is a continuous linear mapping
$\gamma: \CC_0^1(\R^2)\subset \hil^1({\mathcal S'})\to
{\LL}^2(\mathcal S)$,
$\gamma(\varphi(r,\theta))=\varphi(a,\theta)$, that is, there is
$C>0$ so that
\begin{equation}
\|\gamma\varphi\|_{{\LL}^2(\mathcal
S)}=\|\varphi(a,\theta)\|_{{\LL}^2(\mathcal S)}\le
C\,\|\varphi\|_{\hil^1({\mathcal S'})},\quad \varphi\in
\CC_0^1(\R^2).
\end{equation} Note that for $\varphi\in \CC_0^1(\R^2)$ the boundary values
$\varphi(a,\theta)$ are well defined
for any angle $\theta$. By density, this mapping has a unique continuous
extension
$\gamma_0:\hil^1({\mathcal S'})\to {\LL}^2(\mathcal S)$, called the {\em
trace
mapping} (see chapters~1 and~2 of \cite{LM} and
also
\cite{Adams,Brezis}), and one defines
$\psi(a,\theta)\ddd(\gamma_0\psi)(\theta)$ for all
$\psi\in\hil^1({\mathcal S'})$.

Similarly it is defined the trace mapping
\begin{equation}
\gamma_1 : \hil^2(\mathcal S')\to \LL^2(\mathcal S), \qquad
\left.\displaystyle\frac{\partial\psi}{\partial\vec{\mathrm
n}}\right|_{\mathcal S}=\gamma_1\psi ,
\end{equation} where $\vec{\mathrm n}$ is the normalized vector normal to $\mathcal S$
pointing to inside the solenoid.

We shall also make use of the Green's formulae
\begin{equation}
\int_{\mathcal S'}\Delta\psi(x,y)\varphi(x,y)\,dx dy + \int_{\mathcal
S'}\nabla\psi(x,y)\nabla\varphi(x,y)\,dx dy = \int_{\mathcal
S}\gamma_1\psi\;\gamma_0\varphi\,d\sigma,
\end{equation}
which holds for all $\psi,\varphi\in\hil^2(\mathcal S')$, and
\begin{equation}
\int_{\mathcal S'}\frac{\partial\psi}{\partial x}(x,y)\varphi(x,y)\,dx dy
+ \int_{\mathcal S'}\psi(x,y)\frac{\partial\varphi}{\partial x}(x,y)\,dx
dy = \int_{\mathcal S}\gamma_0\psi\;\gamma_0\varphi\;\gamma_0(\vec{\mathrm
n}\cdot\vec{e_x})\,d\sigma,\end{equation}
\begin{equation}\int_{\mathcal S'}\frac{\partial\psi}{\partial y}(x,y)\varphi(x,y)\,dx
dy + \int_{\mathcal S'}\psi(x,y)\frac{\partial\varphi}{\partial
y}(x,y)\,dx dy = \int_{\mathcal
S}\gamma_0\psi\;\gamma_0\varphi\;\gamma_0(\vec{\mathrm
n}\cdot\vec{e_y})\,d\sigma,
\end{equation} which hold for all $\psi,\varphi\in\hil^1(\mathcal S')$, where
$d\sigma$ is the ``surface'' measure in $\mathcal S$ (recall that here
$\mathcal S$ is the circle centered at the origin and radius  $a>0$),
$(x,y)\in\R^2$ and $\vec{e_x}$, $\vec{e_y}$ are the unit vectors along the
axes $x$ and $y$, respectively.

Note that the kernel of the trace operator $\gamma_0$ is the Hilbert space
\begin{equation}
\hil^1_0(\mathcal S')\ddd \left\{\psi\in\hil^1(\mathcal S'):
(\gamma_0\psi)(\theta)=\psi(a,\theta)=0\right\},
\end{equation} which can also be defined as the closure of $\CC_0^\infty(\mathcal S')$
in $\hil^1(\mathcal S')$.

Now we introduce a boundary form $\Gamma$ for the initial AB operator
\eqref{initialABh}, and
companion mappings $\rho_1$ and $\rho_2$ as well. The boundary form of
$H$, for $\psi,\varphi\in\dom H^*$, is
\begin{equation}
\Gamma(\psi,\varphi) \ddd \la H^*\psi,\varphi\ra - \la \psi,H^*\varphi\ra,
\end{equation} and by restricting  to those self-adjoint extensions whose domains are
contained in~$\hil^2(\mathcal S')$, Sobolev traces
can be invoked. Since $\mathrm{div\,} \mathbf A = 0$, the boundary form
of $H$ is found to be given by
\begin{equation}
\begin{split}
\Gamma(\psi,\varphi) & =  \left\langle H^\ast\psi, \varphi\right\rangle -
\left\langle \psi, H^\ast\varphi\right\rangle \\
 & =  \int_{\mathcal S} \left( \overline{\gamma_0\psi}\;\gamma_1\varphi -
\overline{\gamma_1\psi}\;\gamma_0\varphi \right)d\sigma -
2i\int_{\mathcal S} \left( \overline{\gamma_0\psi}\;\gamma_0 ({\bf
A}\cdot\vec{\mathrm n})\;\gamma_0\varphi \right) d\sigma,
\end{split}
\end{equation}
for all $\psi,\varphi\in {\mathcal H}^2(\mathcal S')$.

By passing to polar coordinates $(r,\theta)$, the above boundary form
$\Gamma$ may be rewritten as
\begin{multline}
\Gamma(\psi,\varphi) = a\int_0^{2\pi} \Biggl(
\overline{\psi(a,\theta)}\frac{\partial\varphi}{\partial r}(a,\theta) -
\overline{\frac{\partial\psi}{\partial r}(a,\theta)}\varphi(a,\theta)
\Bigg. \\
\Bigg. - 2 i\, \overline{\psi(a,\theta)} \left({\bf
A}\cdot\vec{r}\right)(a,\theta) \varphi(a,\theta) \Biggr) d\theta,
\end{multline}
for all $\psi,\varphi\in {\mathcal H}^2(\mathcal S')$, with
$\chi(a,\theta)$ and $\displaystyle\frac{\partial\chi}{\partial
r}(a,\theta)$ denoting the traces  $\gamma_0\chi$ and $\gamma_1\chi$,
respectively, for all $\chi\in\hil^2(\mathcal S')$.

Now we introduce the boundary triple $({\bf h}, \rho_1, \rho_2)$, with
${\bf h} = {\rm L}^2(\mathcal S)$, acting in ${\mathcal H}^2(\mathcal S')$
by $\rho_j : {\mathcal H}^2(\mathcal S')\to {\rm L}^2(\mathcal S)$,
$j=1,2,$
\begin{equation}
\begin{split}
\rho_1(\psi) & = \psi(a,\theta) + i \left( \frac{\partial\psi}{\partial
r}(a,\theta) - i ({\bf A}\cdot\vec{r})(a,\theta)\psi(a,\theta)\right), \\
\rho_2(\psi) & = \psi(a,\theta) - i \left( \frac{\partial\psi}{\partial
r}(a,\theta) - i ({\bf A}\cdot\vec{r})(a,\theta)\psi(a,\theta)\right).
\end{split}
\end{equation}
After a short calculation it  follows that
\begin{equation}
\left( 2 i/ a\right) \Gamma(\psi,\varphi) = \left\langle
\rho_1(\psi),\rho_1(\varphi) \right\rangle_{{\rm L}^2(\mathcal S)} -
\left\langle \rho_2(\psi),\rho_2(\varphi) \right\rangle_{{\rm
L}^2(\mathcal S)},
\end{equation}
for all $\psi,\varphi\in {\mathcal H}^2(\mathcal S')$.

Since  ${\bf A}\cdot\vec{r}=0$, the expressions of $\rho_1$ and $\rho_2$
are reduced to
\begin{equation}
\begin{split}
\rho_1(\psi) & = \psi(a,\theta) + i \frac{\partial\psi}{\partial
r}(a,\theta), \\
\rho_2(\psi) & = \psi(a,\theta) - i \frac{\partial\psi}{\partial
r}(a,\theta),
\end{split}
\end{equation}
and the vector potential no longer appears in these expressions; note that
this was possible only due to the cylindrical symmetry of the problem.

Finally, by applying Theorem~\ref{extviatriplas}, the above constructions
permit us to conclude the main result of this section:

\begin{teor2}\label{mainTeor}
All self-adjoint extensions $H^U$ of $H$, acting in ${\mathcal
H}^2(\mathcal S')$, are characterized by unitary operators $U : {\rm
L}^2(\mathcal S)\to {\rm L}^2(\mathcal S)$ so that $\rho_2(\psi) =
U\rho_1(\psi)$, that is,
\begin{equation}
\begin{split}
\dom H^U & = \left\{\psi\in {\mathcal H}^2(\mathcal S') : (\Id -
U)\psi(a,\theta) = i (\Id + U)\frac{\partial\psi}{\partial
r}(a,\theta)\right\}, \\
H^U\psi & = H^\ast\psi,\qquad\psi\in \dom H^U.
\end{split}
\end{equation}
\end{teor2}

\subsection{Some self-adjoint extensions of $H$}
For sake of completeness, in what follows we present some particular
choices of unitary operators $U$ and the corresponding self-adjoint
extensions \cite{ISTQD} of the initial AB operator~$H$.

\begin{exem2}\label{EEAAD}
If $U = -\Id$, then
\begin{equation}
\begin{split}
\dom H^{U} & = \left\{\psi\in {\mathcal H}^2(\mathcal S') :
\psi(a,\theta) = 0 \right\} = {\mathcal H}^2(\mathcal S')\cap {\mathcal
H}^1_0(\mathcal S'), \\
H^{U}\psi & = H^\ast\psi,\qquad \psi\in \dom H^{U}.
\end{split}
\end{equation}
This is the so-called  Dirichlet self-adjoint realization, which is
usually assumed to be the physical relevant in the literature
\cite{Ruij,deOP}.
\end{exem2}

\begin{exem2}
If $U = \Id$, then
\begin{equation}
\begin{split}
\dom H^{U} & = \{\psi\in {\mathcal H}^2(\mathcal S') :
\partial\psi/\partial r(a,\theta) = 0 \}, \\
H^{U}\psi & = H^\ast\psi,\qquad \psi\in \dom H^{U}.
\end{split}
\end{equation}
This is the so-called  Neumann self-adjoint realization.
\end{exem2}

\begin{exem2}
Assume that $(\Id + U)$ is invertible.

In this case, to each self-adjoint operator $A:\dom A\subset
{\LL}^2(\mathcal S)\to
{\LL}^2(\mathcal S)$ corresponds a
self-adjoint extension $H^A$. In fact, first pick a unitary operator~$U_A$
so that $A=-i(\Id-U_A)(\Id+U_A)^{-1}$,  $\dom
A=\img (\Id+U_A)$ and
$\img A=\img (\Id-U_A)$; remind of Cayley transform. Now, $\dom H^A$ is
the set of
$\psi\in\hil^2(\mathcal S')$ with ``$\partial\psi/\partial r(a,\cdot)= A
\psi(a,\cdot)$,''  understood in the sense
that
\begin{equation}
\left( \Id-U_A \right)\psi(a,\theta) = i (\Id+U_A)\frac{\partial
\psi}{\partial r}(a,\theta),
\end{equation} in order to avoid domain questions. Of course the quotation marks may be
removed in case the operator~$A$ is bounded.

Similarly, for each self-adjoint $B$ acting in ${\LL}^2(\mathcal S)$ there
corresponds a unitary $U_B$, and if
$(\Id-U_B)$ is invertible, then it corresponds the self-adjoint
extension~$H^B$ of~$H$ with
$\dom H^B$ being the set of
$\psi\in\hil^2(\mathcal S')$ so that ``$\psi(a,\cdot)=B
\frac{\partial\psi}{\partial r} (a,\cdot)$,'' in the sense that
\begin{equation}
\left( \Id-U_B \right)\psi(a,\theta) = i (\Id+U_B)\frac{\partial
\psi}{\partial r}(a,\theta).
\end{equation} The quotation marks may be removed in case the operator~$B$ is bounded.

Note that Example~\ref{exampMultiplU} ahead is, in fact, particular cases
of this example in which $A=\mathcal
M_f$ and $B=\mathcal M_g$ are multiplication operators.
\end{exem2}

\begin{exem2}\label{exampMultiplU}
$U$ is a multiplication operator.

Given a real-valued (measurable) function $u(\theta)$ defined on
$\mathcal S$, put $U=\mathcal M_{e^{iu(\theta)}}$. If the set $\{\theta:
\exp(iu(\theta))=-1\}$ has measure zero, then the function
\begin{equation}
f(\theta) = -i\frac{1-e^{iu(\theta)}}{1+e^{iu(\theta)}}
\end{equation} is (measurable) well defined and real valued. The domain of the
corresponding self-adjoint extension $H^U$ of~$H$ is
\begin{equation}
\dom H^U = \left\{ \psi\in\hil^2(\mathcal S'): \partial\psi/\partial
r(a,\theta)=f(\theta)\psi(a,\theta) \right\}.
\end{equation} Similarly, if  $\{\theta:
\exp(iu(\theta))=1\}$ has measure zero,
\begin{equation}
g(\theta)=i\frac{1+e^{iu(\theta)}}{1-e^{iu(\theta)}}
\end{equation} is real valued and the domain of the subsequent self-adjoint extension
$H^U$
of~$H$ is
\begin{equation}
\dom H^U = \left\{ \psi\in\hil^2(\mathcal S'):
\psi(a,\theta)=g(\theta)\partial\psi/\partial r(a,\theta)\right \}.
\end{equation} Special cases are given by constant functions $f,g$, the so-called
Robin self-adjoint realization. We shall discuss the scattering for this
extension in Section~\ref{SectionScatter}.
\end{exem2}

\begin{obs2}\label{obsHinfiDI}
Since the deficiency indices of~$H$ are infinite, there is a plethora of
self-adjoint extensions of $H$ in
the multiply connected domain $\mathcal S'$. Some of them can be quite
unusual and hard to
understand from the physical and mathematical points of view.
\end{obs2}

\begin{obs2}\label{remaHneg}
By using a continuous extension of the trace maps to the dual Sobolev spaces
$\hil^{-1/2}(\mathcal S)$ and
$\hil^{-3/2}(\mathcal S)$, in \cite{G06} one finds references and comments
to her previous works on
all self-adjoint extensions of the laplacian in terms of self-adjoint
operators from closed subspaces of
$\hil^{-1/2}(\mathcal S)$. It is possible to follow those works and apply
the same technique to find all self-adjoint extensions of the initial AB
operator $H$, but we will not describe them here since the
characterizations are rather abstract, involve spaces not  usual in
quantum mechanics, and they require a length construction that is not so
clean as the extensions we have found in Theorem~\ref{mainTeor} (which
also includes the traditional self-adjoint extensions we are most
interested in).
\end{obs2}

\section{Scattering}\label{SectionScatter}
In this section we study and compare the scattering for the Robin
self-adjoint
realizations (which includes Neumann and Dirichlet as particular cases) of
the initial AB operator $H$.

\subsection{Scattering theory: a brief account}

Now we briefly recall results from scattering theory, based
mainly on \cite{Ruij, AJS77}, focused on what we want to do in the next
sections. For details and a thorough study of the mathematics of
scattering theory see \cite{RS3, Y}.

Consider a system in (non-relativistic) quantum mechanics whose states
$\xi$ are unit vectors in a Hilbert space $\hil$ and whose time evolution
is generated by a self-adjoint operator $h$ acting in $\hil$, and let
$H_0$ denote the free hamiltonian acting in $\hil_0$. The question is
whether the states $e^{-i h t}\xi$ are scattering states, i.e., if there
are free states $\xi_\pm\in\hil_0$ so that
\begin{equation}
\left\| e^{-i h t}\xi - \mathcal J e^{- i H_0 t}\xi_\pm \right\| = \left\|
\xi - e^{i h t}\mathcal J e^{- i H_0 t}\xi_\pm \right\|
\end{equation}
vanishes as $t\to\pm\infty$. A comparison mapping $\mathcal
J:\hil_0\to\hil$, which is a unitary operator (or just a bounded one), is
sometimes conveniently
introduced.

\begin{defi2}\label{WO}
The {\it wave operators} are the strong limits
\begin{equation}
\mathcal W_\pm\ddd {\mathrm s}\text -\lim_{t\to\pm\infty} e^{i h t}\mathcal
J e^{- i H_0 t},
\end{equation}
if they exist.
\end{defi2}

Recall that the wave operators $\mathcal W_\pm$ are said to be  {\it
complete} if $\img
\mathcal W_\pm = \hil_{{\rm p}}(h)^\bot$, where $\hil_{{\rm p}}(h)$
denotes the closure of the subspace spanned by the eigenvectors of $h$.

The vectors $\xi_\pm$ and $\xi$ satisfy the relation $\xi = \mathcal
W_\pm\xi_\pm$ and  the wave operators $\mathcal W_{\pm}:\dom \mathcal
W_{\pm} \to \img \mathcal W_{\pm}$ are partial isometries. Thus, $\mathcal
W_{\pm} \mathcal W_{\pm}^\ast$ are orthogonal projections onto $\img
\mathcal W_{\pm}$, and restricted to these subspaces $\mathcal
W_{\pm}^\ast = \mathcal W_{\pm}^{-1}$. Furthermore,
\begin{equation}
\xi_+ = S\xi_-,
\end{equation}
where $S \ddd \mathcal W_+^\ast \mathcal W_-$ is the so-called  {\it
scattering operator} or $S$-{\it matrix}.

 The physical system we consider is a scattering of particles off a
cylindrical obstacle in the plane of points $\vec x=(x,y)$, and the
Hilbert space is $\hil=\LL^2( \mathcal S')$. Moreover, we are also
interested in the  scattering away from a short range continuous
potential  $V(x,y)$, which we also assume that it is spherically symmetric,
that is, $V(x,y)=V(r)$, $r=|(x,y)|$; short range means that there are
constants $C,R > 0$ so that
\begin{equation}
| V(x,y)|\le \frac{C}{r^{1+\delta}}, \quad \forall\,r> R,
\end{equation}
for some $\delta > 0$. Let
\begin{equation}
h = - \Delta + V(r)\quad {\rm and}\quad H_0 = p_1^2+p_2^2,
\end{equation}
acting in the position space $\hil$ and momentum $\hil_0 =
\LL^2(\hat{\R}^2)$, respectively, where $\vec{p}=(p_1, p_2)$,
$p^2=|\vec{p}|^2$ and the comparison operator $\mathcal J$ is the inverse
Fourier transform~$\mathcal F$.

In the time-independent scattering theory one solves the time-inde\-pend\-ent
Schr\"odinger equation  $h\varphi = p^2\varphi$ for the incoming
$\varphi_{-}(\vec{x},\vec{p})$ and outgoing $\varphi_+(\vec{x},\vec{p})$
functions,
which are reduced to a plane wave $\phi(\vec{x},\vec{p})= e^{i
\vec{x}\cdot\vec{p}}$ for $|\vec{x}|\to \infty$ (these are solutions of
the Schr\"odinger equation of the free particle). One has the following
connection
\begin{equation}
(\mathcal W_{\pm}\zeta)(\vec{x})= \frac{1}{2\pi}\int
\varphi_{\pm}(\vec{x},\vec{p})\zeta(\vec{p})d\vec{p}, \quad
\zeta\in\hil_0.
\end{equation}

By employing polar coordinates, both in space position $(r,\theta)$ and in
the space of momenta $(k,\theta')$, one obtains
\begin{equation}
h = -\frac{\partial^2}{\partial r^2} - \frac{1}{r}\frac{\partial}{\partial
r} - \frac{1}{r^2}\frac{\partial^2}{\partial \theta^2} + V(r)\quad {\rm
and}\quad H_0 = k^2,
\end{equation}
respectively. One then considers the asymptotic behaviour of the solution
(incoming wave function) of the time-independent Schr\"odinger equation
\begin{equation}\label{ondain}
\varphi_-(r, \theta; k, \theta') \sim e^{i k r \cos(\theta - \theta')} +
f(k, \theta- \theta') \frac{e^{i k r}}{r^{1/2}}, \quad r\to\infty,
\end{equation}
where $f$ is the {\it scattering amplitude}, and so the {\it differential
scattering cross section} is
\begin{equation}
\left( \frac{d\sigma}{d\theta}\right)(k,\theta) = |f(k,\theta)|^2.
\end{equation}
Physically, this quantity measures the probability density of an incident
particle, after interaction with the target (scatterer center), i.e., a
scattered particle to be found within of a cone around $(k,\theta)$.

Now we recall how to find the asymptotic behaviour of $\varphi_-$ and the
scattering amplitude. If $,J_n$ denotes the Bessel function of first kind
of order $n$, one has
\begin{equation}
e^{i k r \cos \theta} = \sum_{m=-\infty}^\infty i^{|m|}J_{|m|}(k r) e^{i m
\theta},\quad f(k, \theta) = \sum_{m=-\infty}^\infty f_m(k) e^{i m
\theta},
\end{equation}
and by \eqref{ondain}, the asymptotic behaviour of $J_n(r)$ \cite{Olver},
\begin{equation}
J_n(r) \sim \left(\frac{2}{\pi r} \right)^{1/2}\cos\left( r - \frac{1}{2}n
\pi - \frac{\pi}{4} \right),\quad r\to\infty,
\end{equation}
and recalling that $\cos\vartheta = (e^{ i \vartheta} + e^{- i
\vartheta})/2$, one obtains
\begin{equation}\label{compassintWFI}
\varphi_-(r,\theta; k, 0)\sim \sum_{m=-\infty}^{\infty} \left[
\frac{(-1)^m e^{i\pi/4 - i k r} }{(2\pi k r)^{1/2}} + \left( \frac{e^{- i
\pi/4}}{(2\pi k r)^{1/2}} + \frac{f_m}{r^{1/2}} \right) e^{i k r}  \right]
e^{i m \theta}.
\end{equation}

On the other hand, $\varphi_-$ solves the Schr\"odinger equation and is
regular at the origin. By using separation of variables,
\begin{equation}\label{expsol}
\varphi_-(r, \theta; k, 0) = \sum_{m=-\infty}^\infty a_m(k) \varphi_m(r,
k) e^{i m \theta},
\end{equation}
where $\varphi_m$ is a solution for radial Schr\"odinger equation of
angular momentum~$m$,
\begin{equation}
\left( -\frac{d^2}{d r^2}-\frac{1}{r}\frac{d}{d r}+ \frac{m^2}{r^2}+
V(r)\right)\varphi = k^2\varphi,
\end{equation}
which is regular at the origin. Since $V$ is short range, it follows that
the behaviour of $\varphi_m$ for $r\to\infty$ is given by
\begin{equation}\label{217}
\varphi_m(r,k)\sim \left(\frac{2}{\pi k r}\right)^{1/2}\cos \left( k r
-\frac{1}{2}|m|\pi -
\frac{\pi}{4}+\delta_m(k) \right),
\end{equation}
and herein the phase shift $\delta_m$ was introduced. The phase shift is a
measure of the argument difference  to the asymptotic behaviour of the
solution  $J_{|m|}(k r)$ to the radial free equation that is regular at
the origin.

The  comparison of the asymptotic behaviour of $\varphi_-$ given by
\eqref{compassintWFI}, with the one given by \eqref{expsol} and
\eqref{217}, gives us the important relations
\begin{equation}
a_m(k) = i^{|m|} e^{i \delta_m(k)},
\end{equation}
\begin{equation}
f_m(k) = \frac{e^{2 i \delta_m(k)} -1}{(2\pi i k )^{1/2}},
\end{equation} and
\begin{equation}\label{ampesp}
f(k, \theta) = \frac{1}{(2\pi i k )^{1/2}}\sum_{m=-\infty}^\infty \left(
e^{2 i \delta_m(k)} -1 \right)e^{i m \theta},
\end{equation}
which formally expresses the scattering amplitude in terms of the phase
shifts $\delta_m$. Later on, in our applications to the extensions of the
AB hamiltonian, these relations will be considered from the point of view
of distribution theory.

Now a bit of the time-dependent approach again. Decompose the Hilbert
spaces $\hil$ and $\hil_0$ in subspaces ${\sf h}_m$, ${\sf h}_{0,m}$, with
corresponding projections $P_m$, $P_{0, m}$, respectively. For example,
\begin{equation}
(P_m \eta)(r,\theta) = \frac{e^{i m \theta}}{(2\pi)^{1/2}} \eta_m(r),
\end{equation}
where
\begin{equation}
\eta_m(r)= \frac{1}{(2\pi)^{1/2}}\int_{0}^{2\pi} e^{- i m \theta} \eta(r,
\theta) d\theta
\end{equation}
are the components of $\eta$ with angular momentum $m$, belonging to the
Hilbert space $\hil_r = \LL^2_{r dr}( [0, \infty) )$. The subspaces ${\sf
h}_m$  are invariant under the hamiltonian $h$ and the subsequent
restriction is given by
\begin{equation}
h_m = -\frac{d^2}{d r^2}-\frac{1}{r}\frac{d}{d r}+ \frac{m^2}{r^2}+ V(r)
\end{equation}
in $\hil_r$. For the inverse Fourier transform $\mathcal F$ one has
\begin{equation}
(\mathcal F \zeta)(r, \theta) = (2\pi)^{-1/2}\sum_{m=-\infty}^\infty e^{i
m \theta} (\mathcal F_m \zeta_m)(r),
\end{equation}
with $\mathcal F_m : \hil_k \to \hil_r$ denoting the unitary operator
\begin{equation}
(\mathcal F_m \psi)(r) = i^{|m|}\int_0^\infty J_{|m|}(k r) \psi(k) k dk,
\end{equation}
where $\hil_k = \LL^2_{k dk}([0, \infty))$. Correspondingly, we consider a
sequence of wave operators from $\hil_k$ into $\hil_r$
\begin{equation}
\mathcal W_{\pm,m} = {\mathrm s}\text-\lim_{t \to \pm\infty} e^{i h_m
t}\mathcal F_m e^{- i k^2 t}.
\end{equation}
Since  $\mathcal W_{\pm,m}$ exist and are complete in each sector $m$, one
has the following relation with the time-independent approach
\begin{equation}
(\mathcal W_{\pm,m} \psi)(r) = i^{|m|} \int_0^\infty \varphi_m (k,
r)e^{\mp i \delta_m(k)} \psi(k) k dk,
\end{equation}
so that the corresponding $S$-matrix in the sector $m$, $S_m : \hil_k \to
\hil_k$,
\begin{equation}
S_m = \mathcal W_{+,m}^\ast\mathcal W_{-,m},
\end{equation}
is given, after some calculations, by
\begin{equation}
(S_m \psi)(k) = \left\{ \begin{array}{ll}
e^{2 i \delta_m(k)} \psi(k), & k=k'\\
0, & k\neq k'
\end{array}\right.,
\end{equation}
that is, $S_m$ is simply the multiplication operator by $e^{2 i
\delta_m(k)}$ on $\hil_k$.

The wave operator in Definition~\ref{WO} can now be written as
\begin{equation}
(\mathcal W_\pm \zeta)(r, \theta) = (2 \pi)^{-1/2}\sum_{m=-\infty}^\infty
e^{i m \theta} (\mathcal W_{\pm,m} \zeta_m)(r).
\end{equation}
Similarly, the scattering operator $S:\hil_0 \to \hil_0$ is found to satisfy
\begin{equation}
\begin{split}
\langle \zeta, S \xi \rangle
= \langle \zeta, \xi \rangle  &+  (2\pi)^{-1}\sum_{m=-\infty}^\infty
\int_0^\infty \int_0^{2\pi} \int_0^{2\pi} e^{i m (\theta - \theta')}
\overline{\zeta(k, \theta)} \\
&\times  \xi(k, \theta')\left( e^{2 i \delta_m(k)} - 1 \right) d\theta
d\theta' \,k\, dk,
\end{split}
\end{equation} and by equation \eqref{ampesp}, we conclude that
\begin{equation}
\begin{split}
\langle \zeta, S \xi \rangle = \langle \zeta, \xi \rangle &+ \int_0^\infty
\int_0^{2\pi} \int_0^{2\pi} \overline{\zeta(k, \theta)} \xi(k, \theta') \\
&\times 
\left( \frac{ik}{2\pi}\right)^{1/2} f(k, \theta - \theta') d\theta
d\theta' \,k\, dk,
\end{split}
\end{equation}
for all $\zeta,\xi\in\hil_0$, where $\langle\cdot, \cdot \rangle$ denotes
the inner product in $\hil_0$, and so
\begin{equation}
``(S - \Id)(k, \theta) = \left( \frac{ik}{2\pi} \right)^{1/2} f(k,\theta),"
\end{equation}
understood in the sense described above, which is the correct relation
between $f$ and $S$.

\subsection{Robin self-adjoint extensions}
In this subsection we describe the self-adjoint extensions of the initial
AB operator \eqref{initialABh} for which we will study scattering. We
choose some of
the self-adjoint
extensions that preserve angular momentum since we are considering that
the subspaces ${\sf h}_m$ are invariants under $H$. In addition, the
extensions described below are the most common and studied in the
literature when borders are considered. Note that there are extensions that
do not preserve angular momentum.

In order to simplify expressions, we will take $c=q=1$ in our initial
hermitian operator \eqref{initialABh},
\begin{equation}
H = (-i\nabla - {\bf A})^2,
\end{equation}
which acts in a subspace of the Hilbert space $\hil = \LL^2(\mathcal S')$,
and recall
that the vector potential ${\bf A}$, in polar coordinates $(r,\theta)$, is
given by ${\bf A} = (A_r,A_\theta)$, with $A_r\equiv 0$ and $A_\theta =
\displaystyle\frac{\Phi}{2\pi r}$, $r \geq a$. This operator can be
written in polar coordinates as
\begin{equation}
H =  -\frac{\partial^2}{\partial r^2} -
\frac{1}{r}\frac{\partial}{\partial r} + \frac{1}{r^2}\left(
i\frac{\partial}{\partial\theta} - \alpha \right)^2,
\end{equation}
where $\alpha = -\Phi/(2\pi)$. Without loss of generality,  consider
$0\leq \alpha<1$, and $\alpha=0$ means that no magnetic field is present.

The next step is to construct the self-adjoint extensions of $H$ we are
interested in. By
making the polar decomposition $\hil = \hil_r^a \otimes \hil_\theta$,
where $\hil_r^a = \LL^2_{r dr}[a, \infty)$ and
$\hil_\theta=\LL^2[0,2\pi]$, we obtain a sequence of formal restriction
operators to~${\sf
h}_m$
\begin{equation}
H_{m+\alpha} = -\frac{d^2}{dr^2} - \frac{1}{r}\frac{d}{dr} +
\frac{(m+\alpha)^2}{r^2}
\end{equation}
in $\hil_r^a$. To achieve our goal we need to turn these operators into
self-adjoint ones acting in $\hil_r^a$. They are not essentially
self-adjoint on $\CC_0^\infty(a, \infty)$ for any $m+\alpha$. To see this,
note that the potential term $(m+\alpha)^2/r^2$ is bounded in $\hil_r^a$,
hence we need only to consider the differential operator
\begin{equation}
\textsl{h}_1 = -\frac{d^2}{dr^2} - \frac{1}{r}\frac{d}{dr} + \frac{1}{4 r^2}.
\end{equation}
But using the unitary operator $U:\hil_r^a \to \LL^2_{dr}[a,\infty)$,
given by $(U \psi)(r)= r^{1/2}\psi(r)$, the operator $\textsl{h}_1$
becomes
\begin{equation}
\textsl{h}_2 =U \textsl{h}_1 U^{-1} = -\frac{d^2}{dr^2},
\end{equation}
which is not essentially self-adjoint on $\CC_0^\infty(a, \infty)$ since
the functions $u_{\pm}(r) = e^{-e^{\pm i\pi/4}r}\in\LL^2_{dr}[a,\infty)$
and satisfy $\textsl{h}_2^\ast u \pm i u =0$; in other words, its
deficiency indices are $n_-(\textsl{h}_2)=n_+(\textsl{h}_2)=1$. Since
this holds for all $m$, it justifies the assertion that
$n_{\pm}(H)=\infty$  in Remark~\ref{obsHinfiDI}.

However, we can find all self-adjoint extensions of $\textsl{h}_2$
\cite{ISTQD}, which are well known and given by
\begin{equation}
\dom \textsl{h}_2^{\tilde\lambda} = \left\{\psi\in\hil^2[a,\infty) :
\psi(a) = \tilde\lambda \psi'(a) \right\}, \quad
\textsl{h}_2^{\tilde\lambda}\psi = \textsl{h}_2\psi,
\end{equation}
for each $\tilde\lambda\in \R \cup \{\infty \}$.

Thus, we have the corresponding self-adjoint extensions of $H_{m+\alpha}$
\begin{equation}
\dom H_{m+\alpha}^{\tilde\lambda} = U^{-1} ( \dom
\textsl{h}_2^{\tilde\lambda} ),\quad H_{m+\alpha}^{\tilde\lambda}\psi =
H_{m+\alpha}\psi,
\end{equation}
that is,
\begin{equation}
\begin{split}
\dom H_{m+\alpha}^{\tilde\lambda} & = \left\{ u = r^{-1/2}\psi :
\psi\in\hil^2[a,\infty)\,\,{\rm and}\,\,  \psi(a) = \tilde\lambda \psi'(a)
\right\} \\
& = \left\{ u\in U^{-1}\left( \hil^2[a,\infty)\right) : r^{1/2}u(r)|_{r=a}
= \tilde\lambda \frac{d}{dr}[r^{1/2}u(r)]|_{r=a} \right\}.
\end{split}
\end{equation}
Therefore, the boundary conditions that characterize the Robin
self-adjoint extensions of $H_{m+\alpha}$ are given by $(2a -
\tilde\lambda) u(a) = 2a\tilde\lambda  u'(a)$. If $\tilde\lambda\neq 2 a$,
then
\begin{equation}
\dom H_{m+\alpha}^{\tilde\lambda} = \left\{ u\in U^{-1}\left(
\hil^2[a,\infty)\right) : u(a) = \frac{2a\tilde\lambda}{2a -
\tilde\lambda} u'(a) \right\}.
\end{equation}

We shall denote these Robin self-adjoint extensions  by
$H_{m+\alpha}^{\lambda}$, that is
\begin{equation}
\begin{split}
\dom H_{m+\alpha}^{\lambda} &= \left\{ u\in U^{-1}\left(
\hil^2[a,\infty)\right) : u(a) = \lambda u'(a) \right\}, \\
H_{m+\alpha}^{\lambda} u &= H_{m+\alpha}u,
\end{split}
\end{equation}
where $\lambda = 2a\tilde\lambda / (2a - \tilde\lambda)$.

Note that upon  integrating by parts it follows that if $\lambda\geq 0$,
then $\left\langle H_{m+\alpha}^{\lambda} u, u\right\rangle\geq 0$, for
all $u\in \dom H_{m+\alpha}^{\lambda}$, that is, the self-adjoint operator
$H_{m+\alpha}^{\lambda}$ is non-negative and, therefore,
$\sigma(H_{m+\alpha}^{\lambda})\subset [0,\infty)$;  from now on we
assume that $\lambda\geq 0$.

Since each sector is invariant under $H$, the Robin self-adjoint extension
of full operator $H$ is given by
\begin{equation}
H^\lambda = \bigoplus_{m\in\Z} H_{m+\alpha}^\lambda \otimes \Id,
\end{equation}
and note that this extension is a special case of
Example~\ref{exampMultiplU}, and the choices of constant functions $f,g$
guarantee that such self-adjoint extensions preserve angular momentum. The
principal cases are for $\lambda = 0$ and
$\lambda = \infty$, which correspond to the well-known self-adjoint
extensions of Dirichlet and Neumann, respectively.

\subsection{Scattering for the Robin extensions}
In this subsection we study the scattering for Robin self-adjoint
realizations of the initial AB operator $H$ introduced in equation
\eqref{initialABh}. We find the scattering
operator $S$, we prove the existence of wave operators and that they are
complete, and we also obtain explicit expressions for them. In addition, we
determined the scattering amplitude and hence the differential scattering
cross section for such extensions.

We underline that there is no magnetic field in case $\alpha=0$, and the
scattering is sole due to the presence of the solenoid; in fact, the
Aharonov-Bohm effect is noticed by comparing the results for $\alpha\ne0$
with this reference case~$\alpha=0$.

\subsubsection{Scattering operator}
Assume initially that the wave operators exist and are complete; under
such conditions, we shall write out the scattering operator. The solution
to
\begin{equation}
\left( -\frac{d^2}{d r^2}-\frac{1}{r}\frac{d}{d r}+
\frac{(m+\alpha)^2}{r^2}\right)\varphi = k^2\varphi,
\end{equation}
with the linear combination $\varphi - \lambda\displaystyle
\frac{d\varphi}{d r}$ vanishing at $r=a$, where $\lambda=
\displaystyle\frac{2a\tilde\lambda}{2a - \tilde\lambda}$, is given by
\begin{multline}
\varphi_m^{\lambda}(k,r) = G_m^{\lambda}(k,a)\left[ \left(
N_{|m+\alpha|}(k a) - \lambda N'_{|m+\alpha|}(k a)\right) J_{|m+\alpha|}(k
r) \right. \\ \left.
- \left( J_{|m+\alpha|}(k a) - \lambda J'_{|m+\alpha|}(k a)\right)
N_{|m+\alpha|}(k r) \right],
\end{multline}
with $G_m^{\lambda}(k,a)$ to be determined by imposing condition
(\ref{217}). By using the asymptotic behaviour of $J_\nu$ and $N_\nu$ for
$r\to\infty$, we obtain
\begin{equation}
\begin{split}
\varphi_m^{\lambda}(k,r) \sim & \left(\frac{2}{\pi k r}\right)^{1/2}
G_m^{\lambda}(k,a) \\
& \times \left[ \cos \left( k r- \frac{1}{2}\left|m+\alpha \right|\pi -
\frac{\pi}{4} \right) \left( N_{|m+\alpha|}(k a) - \lambda
N'_{|m+\alpha|}(k a)\right)\right. \\
& \left. - \,\sin \left( k r- \frac{1}{2}\left|m+\alpha \right|\pi -
\frac{\pi}{4} \right) \left( J_{|m+\alpha|}(k a) - \lambda
J'_{|m+\alpha|}(k a)\right)   \right].
\end{split}
\end{equation}
By comparing the above expression with (\ref{217}) we have
\begin{multline}
G_m^{\lambda}(k,a)\left[ \cos \left( k r- \frac{1}{2}\left|m+\alpha
\right|\pi - \frac{\pi}{4} \right) \left( N_{|m+\alpha|}(k a) - \lambda
N'_{|m+\alpha|}(k a)\right) \right. \\ \left. - \sin \left( k r-
\frac{1}{2}\left|m+\alpha \right|\pi - \frac{\pi}{4} \right) \left(
J_{|m+\alpha|}(k a) - \lambda J'_{|m+\alpha|}(k a)\right) \right]
\end{multline}
$$= \cos \left( k r- \frac{1}{2}\left| m \right|\pi - \frac{\pi}{4} +
\delta_m^{\lambda}(k,\alpha) \right),$$
that is,
\begin{equation}
\cos \left( k r- \frac{1}{2}\left|m+\alpha \right|\pi - \frac{\pi}{4}
+\theta_{\lambda} \right) = \cos \left( k r- \frac{1}{2}\left| m
\right|\pi - \frac{\pi}{4} + \delta_m^{\lambda}(k,\alpha) \right),
\end{equation}
with $\theta_{\lambda}$ so that
\begin{equation}
\cos\theta_{\lambda} = \frac{ N_{|m+\alpha|}(k a) - \lambda
N'_{|m+\alpha|}(k a)}{D},\quad \sin\theta_{\lambda} = \frac{
J_{|m+\alpha|}(k a) - \lambda J'_{|m+\alpha|}(k a)} {D},
\end{equation}
with
\begin{equation}
D = \sqrt{\left( N_{|m+\alpha|}(k a) - \lambda N'_{|m+\alpha|}(k a)
\right)^2 + \left( J_{|m+\alpha|}(k a) - \lambda J'_{|m+\alpha|}(k a)
\right)^2}
\end{equation} and, therefore, (\ref{217}) is satisfied if
\begin{equation}
G_m^{\lambda}(k,a) = \frac{1}{D}.
\end{equation}
Note that $D$ never vanishes. In fact, suppose that $D=0$. Then
$N_{|m+\alpha|}(k a) - \lambda N'_{|m+\alpha|}(k a)=0$ and
$J_{|m+\alpha|}(k a) - \lambda J'_{|m+\alpha|}(k a)=0$. So, it follows
that $J_{|m+\alpha|}(k a) N'_{|m+\alpha|}(k a) - N_{|m+\alpha|}(k a)
J'_{|m+\alpha|}(k a)=0$. But this is a contradiction with the Wronskian
$W_{r=a}[J_{|m+\alpha|}(k r), N_{|m+\alpha|}(k r)]= 2/(\pi a)\neq 0$.

Now, comparing the arguments of the cosines above, it is found that the
phase shift $\delta_m^{\lambda}(k,\alpha)$ is given by
\begin{equation}
\delta_m^{\lambda}(k,\alpha) = \Delta_m(\alpha) + \theta_{\lambda},
\end{equation}
where $\Delta_m(\alpha) =\displaystyle\frac{\pi}{2}\left( |m|-|m+\alpha|
\right)$. Therefore, the scattering operator $S_{\alpha,m}^\lambda :
\hil_k \to \hil_k$ for the Robin self-adjoint extension is
\begin{multline}
S_{\alpha,m}^\lambda=e^{2 i \delta_m^{\lambda}(k,\alpha)} = - e^{2 i
\Delta_m(\alpha)}\\ \times\left[ \frac{\left( J_{|m+\alpha|}(k a) - i
N_{|m+\alpha|}(k a)\right) - \lambda\left( J'_{|m+\alpha|}(k a) - i
N'_{|m+\alpha|}(k a) \right)}{\left(J_{|m+\alpha|}(k a) + i
N_{|m+\alpha|}(ka)\right) - \lambda\left( J'_{|m+\alpha|}(k a) + i
N'_{|m+\alpha|}(k a) \right) } \right],
\end{multline}
that is,
\begin{equation}
S_{\alpha,m}^\lambda = - e^{2 i \Delta_m(\alpha)}\left[ \frac{
H_{|m+\alpha|}^{(2)}(k a) - \lambda H_{|m+\alpha|}^{(2)'}(k a)} {
H_{|m+\alpha|}^{(1)}(k a) - \lambda H_{|m+\alpha|}^{(1)'}(k a)}\right],
\end{equation}
where $H_\nu^{(1),(2)}(x) = J_\nu(x) \pm i N_\nu(x)$ are the Hankel
functions \cite{AS, GR}.

\begin{obs2}
Note that for $\lambda = 0$ we get the Dirichlet case and recover the
expression of the scattering operator found in \cite{Ruij}; and, if we
choose $\lambda = \infty$ we obtain the scattering operator for the
Neumann case, namely,
\begin{equation}
S_{\alpha,m}^\mathcal N = - e^{2 i
\Delta_m(\alpha)}\frac{H_{|m+\alpha|}^{(2)'}(k a)}
{H_{|m+\alpha|}^{(1)'}(k a)}.
\end{equation}
\end{obs2}

\subsubsection{Asymptotic behaviours of the scattering
operator}\label{compassint}
Now we describe  the asymptotic behaviour of the scattering operator for
different self-adjoint extensions for both $ka\to\infty$ and $ka\to 0$,
and also compare the results. This is done from the asymptotic behaviour
of Bessel functions.

We begin with the behaviour for $ka\to\infty$. For this we recall that
\begin{equation}
J_{|m+\alpha|}'(k a) = \left( - J_{|m+\alpha|+1}(k a) k +
\frac{|m+\alpha|}{a} J_{|m+\alpha|}(k a) \right),
\end{equation}
and thus, its behaviour for $ka\to\infty$ is given by
\begin{equation}
\begin{split}
J_{|m+\alpha|}'(k a) & \sim -\left(\frac{2}{\pi ka}\right)^{1/2} \cos
\left( ka - (|m+\alpha|+1)\frac{\pi}{2}-\frac{\pi}{4} \right) k \\
&\quad + \frac{|m+\alpha|}{a} \left(\frac{2}{\pi ka}\right)^{1/2}
\cos\left( ka - |m+\alpha|\frac{\pi}{2}-\frac{\pi}{4} \right);
\end{split}
\end{equation}
and so
\begin{align}
\begin{split}
J_{|m+\alpha|}'(k a) & \sim -\left(\frac{2}{\pi ka}\right)^{1/2} \sin
\left( ka - |m+\alpha|\frac{\pi}{2}-\frac{\pi}{4} \right) k \\
&\quad + \frac{|m+\alpha|}{a} \left(\frac{2}{\pi ka}\right)^{1/2}
\cos\left( ka - |m+\alpha|\frac{\pi}{2}-\frac{\pi}{4} \right).
\end{split}
\end{align}
Similarly,
\begin{equation}
N_{|m+\alpha|}'(k a) = \left( - N_{|m+\alpha|+1}(k a) k +
\frac{|m+\alpha|}{a} N_{|m+\alpha|}(k a) \right),
\end{equation}
and its behaviour for $ka\to\infty$ is given by
\begin{align}\label{opespNn}
\begin{split}
N_{|m+\alpha|}'(k a) & \sim \left(\frac{2}{\pi ka}\right)^{1/2} \cos
\left( ka - |m+\alpha|\frac{\pi}{2}-\frac{\pi}{4} \right) k \\
&\quad + \frac{|m+\alpha|}{a} \left(\frac{2}{\pi ka}\right)^{1/2}
\sin\left( ka - |m+\alpha|\frac{\pi}{2}-\frac{\pi}{4} \right).
\end{split}
\end{align}

Now, by considering $\lambda\neq 0$, the asymptotic behaviour of Bessel
functions and their derivatives given above, we get for $ka\to\infty$
\begin{multline}
S_{\alpha,m}^\lambda \sim (-1)^m e^{-2 i k a + i\pi/2} \\
\times \left[ \frac{(ka)^2\lambda^2 + 2i\lambda(\lambda|m+\alpha|-a ) k a
- \lambda^2 |m+\alpha|^2 + 2\lambda |m+\alpha| a - a^2  }{
(ka)^2\lambda^2+ \lambda^2|m+\alpha|^2 - 2\lambda|m+\alpha|+a^2 }
\right],
\end{multline}
and, since the term in square brackets is approximately 1 in this case, we
have
\begin{equation}
S_{\alpha,m}^\lambda \sim (-1)^m e^{-2 i k a + i\pi/2} ,
\end{equation}
for $k a\to\infty$.
This expression coincides with the Neumann case, i.e., the scattering
operator for the Robin case acts as the Neumann case for large energy,
independently
 of $\lambda$, provided that $\lambda\neq 0$. However, it differs from the
Dirichlet case
\begin{equation}
S_{\alpha,m}^\mathcal D \sim (-1)^m e^{-2 i k a - i\pi/2},
\end{equation}
for $ka\to \infty$.

Summing up, for very large energies the scattering
operator does not distinguish different Robin extensions (i.e.,
$0<\lambda\le\infty$) from the Neumann
case $\lambda=\infty$, but it has a different behaviour from
the Dirichlet case $\lambda=0$. In order to try to understand such
behaviour intuitively, let us informally consider the perhaps simplest
situation, that is, the ``free'' unidimensional reflection from a barrier
at the origin $x=0$ with wavefunction $\psi(x)=A(k)
\sin(kx)+B(k)\cos(kx)$, at
least near the origin; Dirichlet and Neumann boundary conditions impose
that $B=0$ and $A=0$, respectively, whereas Robin condition
$\psi(0)=\lambda\psi'(0)$ imposes the energy relation $|\lambda k|^2 =
|B(k)|^2/|A(k)|^2$, and for large energies $k\to\infty$ one has $|A(k)|\ll
|B(k)|$ and in this region the system behaviour becomes similar to the one
dictated by the Neumann condition.

On the other hand, taking into account that the behaviour of the Bessel
functions for $ka\to 0$  \cite{Olver} are given by
\begin{equation}
J_{|m+\alpha|}(ka)\sim [ (1/2)k a]^{|m+\alpha|} / \Gamma(|m+\alpha|+1),
\end{equation}
and
\begin{equation}
N_{|m+\alpha|}(ka)\sim -\frac{\Gamma(|m+\alpha|)}{\pi [(1/2)k
a]^{|m+\alpha|}},
\end{equation}
we get the following behaviour for the scattering operator
$S_{\alpha,m}^\lambda$ for $ka\to 0$
\begin{align}
\begin{split}
S_{\alpha,m}^\lambda & \sim \cos\beta \frac{d_1^2 - d_2
(ka/2)^{4|m+\alpha|}}{d_1^2 + d_2 (ka/2)^{4|m+\alpha|}} - \sin\beta
\frac{2 d_1 d_3 (ka/2)^{2|m+\alpha|} }{d_1^2 + d_2 (ka/2)^{4|m+\alpha|}}
\\
&\quad + i \left[ \sin\beta \frac{d_1^2 - d_2 (ka/2)^{4|m+\alpha|}}{d_1^2
+ d_2 (ka/2)^{4|m+\alpha|}} + \cos\beta \frac{2 d_1 d_3
(ka/2)^{2|m+\alpha|} }{d_1^2 + d_2 (ka/2)^{4|m+\alpha|}}\right],
\end{split}
\end{align}
with $\beta=\pi(|m|-|m+\alpha|)$ and the coefficients
\begin{align}
\begin{split}
d_1 &\ddd -\Gamma(|m+\alpha|)/\pi - \lambda
[2\Gamma(|m+\alpha|+1)-|m+\alpha|\Gamma(|m+\alpha|)]/(\pi a),
\\
d_2&\ddd \Gamma(|m+\alpha|+1)^{-2}(1+
\lambda^2|m+\alpha|^2/a^2-2\lambda|m+\alpha|/a),
\\
d_3&\ddd \Gamma(|m+\alpha|+1)^{-1}(1-\lambda |m+\alpha|/a),
\end{split}
\end{align}
 are independent of $k$. For $m = 0$ and $\alpha = 0$ we have
\begin{align}
\begin{split}
S_{0,0}^\lambda & \sim \frac{1 - \frac{\pi^2}{4\ln(ka)^2}+ \lambda \left(
\frac{2}{a\ln(ka)} - \frac{\pi^2 (ka/2)^2}{a\ln(ka)^2} \right)   +
\lambda^2 \left( \frac{1 - \pi^2(ka/2)^4}{a^2\ln(ka)^2}\right) }  {1 +
\frac{\pi^2}{4\ln(ka)^2}+ \lambda \left( \frac{2}{a\ln(ka)} + \frac{\pi^2
(ka/2)^2}{a\ln(ka)^2} \right)  +\lambda^2 \left( \frac{1 + \pi^2(ka/2)^4
}{a^2\ln(ka)^2}\right) } \\
& \quad + i \frac{\pi}{\ln(ka)} \frac{ 1 + \lambda \left( \frac{1}{a
\ln(ka)}+ \frac{2(ka/2)^2}{a} \right) + \lambda^2
\frac{2(ka/2)^2}{a^2\ln(ka)}  }{1 + \frac{\pi^2}{4\ln(ka)^2}+ \lambda
\left( \frac{2}{a\ln(ka)} + \frac{\pi^2 (ka/2)^2}{a\ln(ka)^2} \right)
+\lambda^2 \left( \frac{1 + \pi^2(ka/2)^4 }{a^2\ln(ka)^2}\right)  },
\end{split}
\end{align}
for $ka\to 0$.

We observed that, for very small energies $ka\to 0$, the sole scattering
operator is not able to distinguish the Robin, Dirichlet and Neumann
self-adjoint extensions of the initial AB hamiltonian \eqref{initialABh}.
Furthermore,  this
occurs both for the case with field ($\alpha\neq 0$) and for the reference
case
without field ($\alpha=0$). These behaviours were numerically recovered.

\subsubsection{Wave operators}
Now, we prove that in fact the wave operators for the Robin self-adjoint
extensions exist and are complete. In addition, we obtain an explicit
expression for them. We begin with a lemma that will be used in the proof
of Theorem~\ref{opondaR}.

\begin{lema2}\label{lemaGU}
Let $A$ and $B$ be self-adjoint operators and $U$ a bounded operator so
that $AU = UB$, assuming that the compositions are well defined. Then
\begin{equation}
e^{-i A t}U = U e^{-i B t}.
\end{equation}
\end{lema2}
\begin{proof}
For each $\xi\in\dom B$ so that $U\xi\in\dom A$, let $u(t)= U e^{-i B
t}\xi$ and $v(t)= e^{-i A t}U\xi$, for all $t\in\R$. We want to show that
$u(t)=v(t)$, for all $t\in\R$. If $w(t)=\left\|v(t) - u(t) \right\|^2$,
then
\begin{align}
\begin{split}
\frac{dw}{dt} & =\frac{d}{dt}\langle v(t)-u(t),v(t)-u(t)\rangle = 2{\rm
Re}\left\langle v(t)-u(t),\frac{d}{dt}[v(t)-u(t)]\right\rangle \\
& = 2{\rm Re}\left[ (-i) \left( \left\langle
e^{-iAt}U\xi,Ae^{-iAt}U\xi\right\rangle + \left\langle U e^{-iBt}\xi, A U
e^{-iBt}\xi\right\rangle\right.\right. \\
& \quad - \left.\left. 2 {\rm Re} \left\langle A e^{-iAt}U\xi,U
e^{-iBt}\xi\right\rangle\right)\right] =0
\end{split}
\end{align}
because what is in square brackets is real since $A$ is self-adjoint.
Therefore $w(t)$ is constant. Since $w(0)=0$, it follows that $w(t)=0$,
for all $t\in \R$.
\end{proof}

Let $P_a : \hil_r \to \hil_r^a$ be given by $(P_a \psi)(r) = ( \chi_{[a,
\infty)}\psi )(r)$, that is, $P_a$ is the orthogonal projection operator
onto $\hil_r^a$.

\begin{teor2}\label{opondaR}
The wave operators
\begin{equation}
\mathcal W_{\pm,\alpha,m}^{\lambda} =  {\mathrm
s\text-}\lim_{t\to\pm\infty} e^{i H_{m+\alpha}^{\lambda} t}P_a \mathcal
F_m e^{- i k^2 t}
\end{equation}
exist and are surjective isometries from $\hil_k$ onto $\hil_r^a$.
Explicitly, for $\psi\in\hil_k$, they are given by the expressions
\begin{equation}\label{caraooR}
\left(\mathcal W_{\pm,\alpha,m}^{\lambda} \psi \right)(r) = i^{|m|}
\lim_{R\to\infty} \int_0^R \varphi_m^{\lambda}(k,r) e^{\mp i
\delta_m^{\lambda}(k,\alpha)}\psi(k) k dk.
\end{equation}
\end{teor2}
\begin{proof}
We consider only the case $\mathcal W_{-,\alpha,m}^{\lambda}$; the proof
for $\mathcal W_{+,\alpha,m}^{\lambda}$ is similar. Since the proof is
rather long, we divide it in three steps:\newline
{\it $1^{{st}}$ Step:} Define the candidate for the limit operator, and
show two equalities. \newline
{\it $2^{ nd}$ Step:} Show that the wave operator $\mathcal
W_{-,\alpha,m}^{\lambda}$ exists and satisfies
(\ref{caraooR}). \newline
{\it $3^{ rd}$ Step:} Show that the wave operator $\mathcal
W_{-,\alpha,m}^\lambda$ is a surjective isometry.

\

{\it $1^{ st}$ Step:} ``Define the candidate for the limit operator, and
show two equalities.''
Let us define an operator $U_{-,\alpha,m}^{\lambda}$ by
\begin{equation}
\left(U_{-,\alpha,m}^{\lambda} \psi \right)(r) = i^{|m|}
\int_\varepsilon^R \varphi_m^{\lambda}(k,r) e^{i
\delta_m^{\lambda}(k,\alpha)}\psi(k) k dk,
\end{equation}
with $\psi\in\hil_k$ and ${\rm supp\,}\psi\subset (\varepsilon, R)$. This
operator is well defined by H\"older's inequality. By using that
\begin{equation}
J_{|\nu|}(kr)=(2/\pi)^{1/2}\cos\left(kr - |\nu|\pi/2 - \pi/4)/(kr)^{1/2} +
O((kr)^{-3/2}\right)
\end{equation}
and
\begin{equation}
N_{|\nu|}(kr)=(2/\pi)^{1/2}\sin\left(kr - |\nu|\pi/2 - \pi/4)/(kr)^{1/2} +
O((kr)^{-3/2}\right),
\end{equation}
it is found that $U_{-,\alpha,m}^{\lambda} \psi \in\hil_r^a$, and it
satisfies
\begin{equation}
\left\| U_{-,\alpha,m}^{\lambda} \psi \right\|_{\hil_r^a}\leq
K_{\lambda}(R,\varepsilon)\left\| \psi \right\|_{\hil_k},
\end{equation}
where $K_{\lambda}(R,\varepsilon)$ depends only on $R$ and $\varepsilon$.

Now we check that $U_{-,\alpha,m}^{\lambda} \psi$ is actually in the
domain of $H_{m+\alpha}^{\lambda}$ and
\begin{equation}
H_{m+\alpha}^{\lambda} U_{-,\alpha,m}^{\lambda} \psi =
U_{-,\alpha,m}^{\lambda} k^2 \psi.
\end{equation}
In fact, let $u\in \dom H_{m+\alpha}^{\lambda}$. Then, by Fubini's
theorem, we can write\newline
\begin{equation}
\left\langle H_{m+\alpha}^{\lambda} u, U_{-,\alpha,m}^{\lambda} \psi
\right\rangle = \int_a^\infty (\overline{H_{m+\alpha}^{\lambda} u})(r)
(U_{-,\alpha,m}^{\lambda} \psi)(r) r dr
\end{equation}
\begin{multline}
= \int_a^\infty \left[
\left(-\frac{d^2}{dr^2}-\frac{1}{r}\frac{d}{dr}+\frac{(m+\alpha)^2}{r^2}
\right)\overline{u}(r)\, i^{|m|} \right. \\
\left. \times \int_\varepsilon^R \varphi_m^{\lambda}(k,r)
e^{i\delta_m^{\lambda}(k,\alpha)} \psi(k) k dk \right] r dr
\end{multline}
\begin{multline}
= i^{|m|}\int_\varepsilon^R e^{i\delta_m^{\lambda}(k,\alpha)} \psi(k)
\left[ (m+\alpha)^2\int_a^\infty \frac{1}{r^2} \varphi_m^{\lambda}(k,r)
\overline u(r) r dr \right. \\
\left. - \int_a^\infty \varphi_m^{\lambda}(k,r) \left(
\frac{d^2}{dr^2}+\frac{1}{r}\frac{d}{dr} \right)\overline u(r) r dr\right]
k dk,
\end{multline}
and integrating by parts the second term in square brackets, we obtain
\begin{multline}
\left\langle H_{m+\alpha}^{\lambda} u, U_{-,\alpha,m}^{\lambda} \psi
\right\rangle = i^{|m|}\int_\varepsilon^R
e^{i\delta_m^{\lambda}(k,\alpha)} \psi(k) \\ \times \left[\int_a^\infty
\left( -\frac{d^2}{dr^2} - \frac{1}{r}\frac{d}{dr} +
\frac{(m+\alpha)^2}{r^2} \right) \varphi_m^{\lambda}(k,r) \overline u(r) r
dr \right] k dk;
\end{multline}
note that there are no boundary terms since
$\varphi_m^{\lambda}(k,a)=\lambda \displaystyle\frac{d
\varphi_m^{\lambda}}{dr}(k,a)$ and $u(a)=\lambda u'(a)$. Therefore,
\begin{equation}
\left\langle H_{m+\alpha}^{\lambda} u, U_{-,\alpha,m}^{\lambda} \psi
\right\rangle = i^{|m|}\int_\varepsilon^R
e^{i\delta_m^{\lambda}(k,\alpha)} \psi(k) \left[\int_a^\infty k^2
\varphi_m^{\lambda}(k,r) \overline u(r) r dr \right] k dk,
\end{equation}
and, again by Fubini, we find that
\begin{equation}
\left\langle H_{m+\alpha}^{\lambda} u, U_{-,\alpha,m}^{\lambda} \psi
\right\rangle = \int_a^\infty \overline u(r)\, i^{|m|}\int_\varepsilon^R
\varphi_m^{\lambda}(k,r) e^{i\delta_m^{\lambda}(k,\alpha)} k^2 \psi(k) k
dk \, r dr,
\end{equation}
that is,
\begin{equation}
\left\langle H_{m+\alpha}^{\lambda} u, U_{-,\alpha,m}^{\lambda} \psi
\right\rangle = \left\langle u, U_{-,\alpha,m}^{\lambda} k^2 \psi
\right\rangle,
\end{equation}
where $\langle \cdot, \cdot \rangle$ denotes the inner product in
$\hil_r^a$. Then $U_{-,\alpha,m}^{\lambda} \psi\in \dom
H_{m+\alpha}^{\lambda} $ and
\begin{equation}
H_{m+\alpha}^{\lambda} U_{-,\alpha,m}^{\lambda} \psi =
U_{-,\alpha,m}^{\lambda} k^2 \psi.
\end{equation}

Finally, apply Lemma~\ref{lemaGU} to conclude
\begin{equation}
e^{- i H_{m+\alpha}^{\lambda} t} U_{-,\alpha,m}^{\lambda} \psi =
U_{-,\alpha,m}^{\lambda} e^{- i k^2 t} \psi.
\end{equation}

\

{\it $2^{ nd}$ Step:} ``Show that the wave operator exists and satisfies
(\ref{caraooR}).'' Assume that $\psi\in \CC_0^\infty(\varepsilon, R)$. By
using the conclusion of the first step (i.e., the last equality above),
one can write,
\begin{equation}
\left\| e^{i H_{m+\alpha}^{\lambda} t}P_a \mathcal F_m e^{- i k^2 t} \psi
- U_{-,\alpha,m}^{\lambda} \psi \right\|^2 = \left\| P_a \mathcal F_m e^{-
i k^2 t} \psi - e^{- i H_{m+\alpha}^{\lambda} t} U_{-,\alpha,m}^{\lambda}
\psi \right\|^2
\end{equation}
\begin{equation}
= \left\| P_a \mathcal F_m e^{- i k^2 t} \psi - U_{-,\alpha,m}^{\lambda}
e^{- i k^2 t} \psi \right\|^2
\end{equation}
\begin{equation}
= \int_a^\infty \left| \left(( P_a \mathcal F_m - U_{-,\alpha,m}^{\lambda}
) e^{- i k^2 t} \psi \right)(r)  \right|^2 r dr
\end{equation}
\begin{multline}
= \int_a^\infty \left| i^{|m|} \int_0^\infty J_{|m|}(k r) e^{- i k^2 t}
\psi(k) k dk \right. \\
\left. - i^{|m|} \int_\varepsilon^R dk\,k \varphi_m^{\lambda} (k, r) e^{i
\delta_m^{\lambda}(k,\alpha)} e^{- i k^2 t} \psi(k) k dk  \right|^2 r dr
\end{multline}
\begin{equation}
= \int_a^\infty \left| i^{|m|} \int_\varepsilon^R e^{- i k^2 t} \psi(k)
\left( J_{|m|}(k r) - \varphi_m^{\lambda} (k, r) e^{i
\delta_m^{\lambda}(k,\alpha)}\right) k dk \right|^2 r dr,
\end{equation}
and after some calculations with the asymptotic behaviour of the two
functions in brackets above, we obtain
\begin{multline}
\left\| e^{i H_{m+\alpha}^{\lambda} t}P_a \mathcal F_m e^{- i k^2 t} \psi
- U_{-,\alpha,m}^{\lambda} \psi \right\|^2 \\
= \int_a^\infty \left| \int_\varepsilon^R e^{- i k^2 t} \psi(k) \left[
K_1^{\lambda}(k a)\frac{e^{i k r}}{(k r)^{1/2}} + K_2^{\lambda}(k a)
O\left( (k r)^{-3/2} \right)\right] k dk \right|^2 r dr,
\end{multline}
with $K_1^{\lambda}$ and $K_2^{\lambda}$ are functions of
class~$\CC^\infty$. Finally, using the inequality $\| f+g\|^2\leq 2\|f\|^2
+ 2\|g\|^2$, we get
\begin{align}
\begin{split}
& \left\| e^{i H_{m+\alpha}^{\lambda} t}P_a \mathcal F_m e^{- i k^2 t}
\psi - U_{-,\alpha,m}^{\lambda} \psi \right\|^2 \\
& \leq 2 \int_a^\infty \left| \int_\varepsilon^R e^{- i k^2 t} \psi(k)
K_1^{\lambda}(k a)\frac{e^{i k r}}{(k r)^{1/2}} k dk \right|^2 r dr \\
&\quad + 2 \int_a^\infty \left| \int_\varepsilon^R e^{- i k^2 t} \psi(k)
K_2^{\lambda}(k a) O\left( (k r)^{-3/2} \right) k dk \right|^2 r dr.
\end{split}
\end{align}

We discuss each term on the right side of the last inequality separately.
By replacing $e^{- i k^2 t + i k r}$ with $(-2 i k t + i r)^{-1}\partial_k
e^{- i k^2 t + i k r}$,  integrating by parts and using the dominated
convergence theorem to  estimate the first term, it is found that it
vanishes as $t\to -\infty$.

For the second term, let $h(k r)= O\left( (k r)^{-3/2} \right)$, then $h(k
r)= M(k r)$ $\times(k r)^{-3/2}$, with $M(k r)$ a bounded function. So, by
Riemann-Lebesgue lemma and dominated convergence theorem,
\begin{multline}
\int_a^\infty \left| \int_\varepsilon^R e^{- i k^2 t} \psi(k)
K_2^{\lambda}(k a) O\left( (k r)^{-3/2} \right) k dk \right|^2 r dr \\
= \int_a^\infty \left| \int_\varepsilon^R e^{- i k^2 t} \psi(k)
K_2^{\lambda}(k a) M(k r) k^{-1/2} dk  \right|^2 r^{-2} dr \to 0,
\end{multline}
as $t\to-\infty$. Then
\begin{equation}
\left\| e^{i H_{m+\alpha}^{\lambda} t}P_a \mathcal F_m e^{- i k^2 t} \psi
- U_{-,\alpha,m}^{\lambda} \psi \right\| \to 0,
\end{equation}
as $t\to -\infty$, and since $\CC_0^\infty(0, \infty)$ is dense in
$\hil_k$, it follows that the wave operator $\mathcal
W_{-,\alpha,m}^{\lambda}$ exists and satisfies (\ref{caraooR}).

\

{\it $3^{ {rd}}$ Step:} ``Show that the wave operator $\mathcal
W_{-,\alpha,m}^\lambda$ is a surjective isometry.'' To show that the wave
operator is an isometry we take $\psi\in\hil_k$ with compact support and
check
\begin{equation}
\left\| \mathcal W_{-,\alpha,m}^{\lambda}\psi \right\| = \lim_{t\to
-\infty} \left\| P_a \mathcal F_m e^{- i k^2 t} \psi \right\| = \lim_{t\to
-\infty}\left\| \mathcal F_m e^{- i k^2 t} \psi \right\| = \left\| \psi
\right\|.
\end{equation}

To prove that $\img \mathcal W_{-,\alpha,m}^\lambda = \hil_r^a$ it
suffices to show that its adjoint is an isometry, because the kernel $\{ 0
\} = \mathrm{N} \left( ( \mathcal W_{-,\alpha,m}^\lambda
)^\ast\right)=\left( \img \mathcal W_{-,\alpha,m}^{\lambda} \right)^\bot
$, and so $\img \mathcal W_{-,\alpha,m}^{\lambda} = \hil_r^a$. Since
$\mathcal W_{-,\alpha,m}^\lambda(\mathcal W_{-,\alpha,m}^\lambda)^\ast$ is
the orthogonal projection onto $\img \mathcal W_{-,\alpha,m}^\lambda$,
which is closed, and since
\begin{align}
\begin{split}
& \left[\mathcal W_{-,\alpha,m}^{\lambda} \left( (\mathcal
W_{-,\alpha,m}^{\lambda})^\ast \psi\right) (k)\right] (r) \\
& = i^{|m|}\lim_{R\to\infty} \int_0^R \varphi_m^{\lambda}(k,r) e^{i
\delta_m^{\lambda}(k,\alpha)}\left( (\mathcal
W_{-,\alpha,m}^{\lambda})^\ast \psi\right) (k) k dk \\
& = i^{|m|}\lim_{R\to\infty} \int_0^R \varphi_m^{\lambda}(k,r) e^{i
\delta_m^{\lambda}(k,\alpha)}\left( (-i)^{|m|}\int_a^\infty
\varphi_m^{\lambda}(k,s) e^{-i \delta_m^{\lambda}(k,\alpha)} \psi(s) s ds
\right) k dk \\
& = \lim_{R\to\infty} \int_0^R G_m^{\lambda}(k,a) D_\nu^{\lambda}(k a, k
r) \int_a^\infty G_m^{\lambda}(k,a) D_\nu^{\lambda}(k a, k s) \psi(s) s
ds\, k dk,
\end{split}
\end{align}
and recalling that $G_m^{\lambda}(k,a) = \displaystyle\frac{1}{D}$, with
\begin{equation}
D = \sqrt{\left( N_{\nu}(k a) - \lambda N'_{\nu}(k a) \right)^2 + \left(
J_{\nu}(k a) - \lambda J'_{\nu}(k a) \right)^2},
\end{equation}
and $\nu=|m+\alpha|$, one has
\begin{equation}
D_\nu^{\lambda} (k a, y) \ddd \left[ N_\nu(k a) - \lambda N'_\nu(k
a)\right] J_\nu(y) - \left[ J_\nu(k a) - \lambda J'_\nu(k a)\right]
N_\nu(y),
\end{equation}
and, in order to conclude the theorem, it is enough to prove the following
lemma.

\begin{lema2}\label{exprFR}
Let $\psi\in \CC_0^\infty(a,\infty)$ and $\nu \geq 0$. Then
\begin{equation}\label{funcaoFR}
\psi(r) = \lim_{R\to\infty}\int_{1/R}^R \frac{1}{D^2} D_\nu^{\lambda} (k
a, k r) \int_a^\infty D_\nu^{\lambda}(k a, k s) \psi(s) s ds\, k dk.
\end{equation}
\end{lema2}

In fact, this lemma implies that
\begin{equation}
\left[\mathcal W_{-,\alpha,m}^{\lambda} \left( (\mathcal
W_{-,\alpha,m}^{\lambda})^\ast \psi\right) (k)\right] (r) = \psi(r),
\end{equation}
for all $\psi\in \CC_0^\infty(a,\infty)$, and since this set is dense in
$\hil_r^a$, it follows that $\img \mathcal
W_{-,\alpha,m}^{\lambda}=\hil_r^a$, and the theorem is proved.
\end{proof}

\

In the following we present the proof of Lemma \ref{exprFR}.
\begin{proof}
Since the Wronskian of $J_\nu(z)$ and $N_\nu(z)$ is equal to $2/(\pi z)$
\cite{Olver}, that is, $W_z [J_\nu, N_\nu] = 2/(\pi z)$, one has
$W_{r}[J_\nu(k r), N_\nu(k r)]= 2/(\pi r)$. Now, we consider the boundary
value  problem
\begin{align}\label{pvfR}
\begin{split}
& (E - H_\nu )\varphi = \psi, \quad a < r < \infty, \quad |{\rm Im\,}E|>0,\\
& \varphi(a)-\lambda\varphi'(a)=0.
\end{split}
\end{align}
The Green's function
\begin{equation}
g(r|s) = \left\{
\begin{array}{ll}
\displaystyle\frac{u_1(r) u_2(s)}{W_s[u_1, u_2]}, & a < r < s, \\
\displaystyle\frac{u_1(s) u_2(r)}{W_s[u_1, u_2]}, & s < r <\infty,
\end{array}
\right.
\end{equation}
is the solution to the auxiliary problem
\begin{align}
\begin{split}
& (E - H_\nu )g = \delta(r-s), \quad a < r < \infty, \\
& g(a)-\lambda g'(a)=0,
\end{split}
\end{align}
where
\begin{align}
\begin{split}
u_1(r) & = \left[ N_\nu(E^{1/2} a) - \lambda N'_\nu(E^{1/2} a)\right]
J_\nu(E^{1/2} r) \\
&\quad - \left[ J_\nu(E^{1/2} a) - \lambda J'_\nu(E^{1/2}
a)\right] N_\nu(E^{1/2} r) \\
& = D_\nu^{\lambda}(E^{1/2} a, E^{1/2} r),
\end{split}
\end{align}
is the solution to $(E - H_\nu )u = 0$ that satisfies the boundary
condition at $r=a$, and
\begin{equation}
u_2(r) = H_\nu^{(1),(2)}(E^{1/2} r),
\end{equation}
is the solution to $(E - H_\nu )u = 0$ that satisfies  the boundary
condition at $\infty$, and the superscripts $(1)$ and $(2)$ correspond to
${\rm Im\,}E>0$ (with ${\rm Im\,}\sqrt{E}>0$) and ${\rm Im\,}E<0$ (with
${\rm Im\,}\sqrt{E}<0$), respectively; $W_s[u_1, u_2]$ is the wronskian of
the solutions $u_1$ and $u_2$ at the point $r=s$, and in this case one has
\begin{equation}
W_s[u_1, u_2] = \frac{2}{\pi s}\left( u_2(a) - \lambda \frac{d u_2}{d
r}(a)\right).
\end{equation}

Write $R_E\ddd (E - H_\nu)^{-1}$ for the resolvent of $H_\nu$ at
``energy'' $E$, so that the solution $(R_E \psi)(r)$ to  problem
(\ref{pvfR}) is given by
\begin{align}
\begin{split}
(R_E \psi)(r) & = \int_a^\infty g(r|s) \psi(s) ds \\
& = \int_a^r \frac{u_1(s) u_2(r)}{W_s[u_1, u_2]} \psi(s) ds +
\int_r^\infty \frac{u_1(r) u_2(s)}{W_s[u_1, u_2]} \psi(s) ds  \\
& = \frac{\pi}{2} \left[ u_2(a) - \lambda \frac{d u_2}{d r}(a)
\right]^{-1} \left[ H_\nu^{(1),(2)}(E^{1/2}r) \int_a^r D_\nu^{\lambda}
(E^{1/2} a, E^{1/2} s)\right. \\
&\quad \left.\times\,\psi(s) s\, ds + D_\nu^{\lambda}(E^{1/2} a, E^{1/2} r)
\int_r^\infty H_\nu^{(1),(2)}(E^{1/2}s) \psi(s) s\, ds \right].
\end{split}
\end{align}

Recall now the Stone formula \cite{ISTQD} for the spectral projection of
$H_\nu$ onto the interval $[a, b]$,
\begin{equation}
\chi_{[a,b]}(H_\nu) = {\mathrm s}\text-\lim_{\delta\to 0^+} \frac{1}{2\pi
i}\int_a^b
(R_{x - i\delta}-R_{x+i\delta}) dx.
\end{equation}
If one writes $E_- = x - i \delta$ and $E_+ = x + i \delta$ for the
``energy'' $E$ with ${\rm Im\,}E<0$ and ${\rm Im\,}E>0$, respectively,
then
\begin{equation}
\left[ \chi_{[1/R,R]}(H_\nu)\psi \right](r) = \lim_{\delta\to 0^+}
\frac{1}{2\pi i}\int_{1/R}^R [(R_{E_-}\psi)(r) - (R_{E_+}\psi)(r)]\, dx.
\end{equation}
Now, substitute the above expressions  for the resolvent operators and use
the dominated convergence theorem (take into account that the functions
$J_\nu$ and $N_\nu$ are continuous and bounded), after some manipulations
and simplifications we obtain the expression
\[
\left[ \chi_{[1/R,R]}(H_\nu)\psi \right](r)
\]
\begin{multline}
= \frac{1}{2}\int_{1/R}^R \frac{D_\nu^{\lambda}(x^{1/2} a,x^{1/2}
r)}{\left( N_\nu(x^{1/2} a) - \lambda N'_\nu(x^{1/2} a) \right)^2 + \left(
J_\nu(x^{1/2} a) - \lambda J'_\nu(x^{1/2} a) \right)^2} \\
\times \int_a^\infty D_\nu^{\lambda}(x^{1/2} a,x^{1/2} s) \psi(s) s\, ds dx.
\end{multline}

Finally, use the change of variable $x^{1/2} = k$, so that $dx /2 =
k\,dk$, to get
\begin{equation}
\left[ \chi_{[1/R,R]}(H_\nu)\psi\right](r) = \int_{1/R}^R  \frac{1}{D^2}
D_\nu^{\lambda}(k a,k r) \int_a^\infty D_\nu^{\lambda}(k a,k s) \psi(s)
s\, ds\,k\,dk,
\end{equation}
which is a.e.\ equal   to the function defined by the integral on the
right side of  (\ref{funcaoFR}). On the other hand,
$\chi_{[a,b]}(H_\nu)\equiv 0$ for $[a, b]\subset (-\infty, 0)$ since
$H_\nu$ is a positive operator (see Theorem 8.3.13 in \cite{ISTQD}), and
so $\sigma(H_\nu)\subset [0,\infty)$ and $H_\nu$ has no
eigenvalues.
\end{proof}

\subsubsection{Scattering amplitude and cross section}
In this subsection we calculate the scattering amplitude and differential
scattering cross section for the Robin self-adjoint extensions, and some
comparisons will be made in the next subsection.

However, first we recall  what was done in \cite{Ruij} to determine the
scattering amplitude $f_\alpha$ for the case of a solenoid of radius zero,
and with Dirichlet condition at the origin; since we will make use of such
results.  In the
case of radius zero, in each sector of angular momentum $m$, one has
\begin{equation}
\Delta_m(\alpha)= \frac{\pi}{2}(|m|-|m+\alpha|),
\end{equation}
for the phase shift, which is a function only of $\alpha$, and so the
corresponding scattering operator is
\begin{equation}
e^{2 i \Delta_m(\alpha)} = \left\{ \begin{array}{ll}
e^{-i \pi \alpha}, & m\geq -\alpha\\
e^{i \pi \alpha}, & m\leq -\alpha
\end{array}\right..
\end{equation}
Then, the Fourier coefficients of $f_\alpha$ in the expression
\eqref{ampesp} has constant modulus and do not vanish as  $|m|\to
\infty$; so the scattering amplitude $f_\alpha$ is seen as a
distribution. To obtain the correct expression of the amplitude
$f_\alpha$, note that the scattering operator $S_\alpha$ on $\hil_0$ is an
integral operator and, by using the above expressions, in \cite{Ruij} it
was found that
\begin{equation}
(S_\alpha \xi)(k,\theta)=\int_{0}^{2\pi} s_\alpha(\theta-\theta')
\xi(k,\theta') d\theta',
\end{equation}
with
\begin{equation}
s_\alpha(\theta) = \delta(\theta) \cos(\pi\alpha) + i
\frac{\sin(\pi\alpha)}{\pi} \mathrm{PV}\left(\frac{1}{e^{i\theta}-1}\right),
\end{equation}
where $\mathrm{PV}$ denotes the principal value (recall that here
$0\le\alpha<1$). These expressions and the relation $(S - \Id)(k, \theta)
= \left( \frac{ik}{2\pi} \right)^{1/2} f(k,\theta)$ imply the following
expression for the scattering amplitude
\begin{equation}\label{falphaComDelta}
f_\alpha(k,\theta)= \left( \frac{2\pi}{ik} \right)^{1/2}\left[
\delta(\theta) [ \cos(\pi\alpha)-1] + i \frac{\sin(\pi\alpha)}{\pi}
\mathrm{PV}\left(\frac{1}{e^{i\theta}-1}\right) \right].
\end{equation}
Now, if $\theta\neq 0$ the distribution $f_\alpha$ is represented by the
function \cite{Ruij}
\begin{equation}\label{exprsFalpha}
f_\alpha(k,\theta)= \frac{\sin(\pi\alpha)}{(2\pi i k)^{1/2}} \frac{e^{-
i\theta/2}}{\sin(\theta/2)},
\end{equation}
and so the differential scattering cross section in this case is
\begin{equation}\label{scr0}
\left( \frac{d\sigma}{d\theta} \right)_\alpha(k,\theta)= \frac{1}{2\pi k}
\frac{\sin^2(\pi\alpha)}{\sin^2(\theta/2)},\quad \theta\neq 0,
\end{equation}
which agree with the expressions found by Aharonov and Bohm \cite{AB59}
and also by other authors, for example in \cite{Hagen}. Thus we will
continue looking at the scattering amplitude as a distribution, which will
be conveniently calculated from the Fourier series.

\begin{obs2}
We observe that in \cite{Hagen} it is advocated that there should be no
$\delta(\theta)$ in the above expression \eqref{falphaComDelta} for the scattering amplitude, and that $f_\alpha$ should be restricted to \eqref{exprsFalpha}; this causes a controversy with references \cite{Hagen} and \cite{Ruij}.
In any event, since we do not consider the forward direction $\theta=0$ in
our comparisons of the scattering due to different self-adjoint
extensions (i.e., our main goal in the next section), we are able to keep away from such controversy.
\end{obs2}

Now we turn to our Robin extensions and positive radius. For $\alpha=0$,
that is, no magnetic
field, the expression \eqref{ampesp} gives for scattering amplitude
associated with $H^\lambda$
\begin{equation}
f_0^\lambda (k, \theta) = \frac{1}{(2\pi i k
)^{1/2}}\sum_{m=-\infty}^\infty \left( e^{2 i \delta_m^\lambda(k,0)} -1
\right)e^{i m \theta},
\end{equation}
and since
\begin{equation}
e^{2 i \delta_m^{\lambda}(k,0)} = - \frac{ H_{|m|}^{(2)}(k a) - \lambda
H_{|m|}^{(2)'}(k a)} { H_{|m|}^{(1)}(k a) - \lambda H_{|m|}^{(1)'}(k a)},
\end{equation}
we obtain
\begin{equation}
f_0^\lambda (k, \theta) = - \left(\frac{2}{\pi i
k}\right)^{1/2}\sum_{m=-\infty}^\infty \frac{J_{|m|}(k a) - \lambda
J'_{|m|}(k a)} { H_{|m|}^{(1)}(k a) - \lambda H_{|m|}^{(1)'}(k a)}e^{i m
\theta}.
\end{equation}
Note that for fixed $ka$ and $k\neq 0$, the series above is convergent
since its coefficients are fast decaying as $|m|\to\infty$, due to the
well-known behaviour of  Bessel functions. Thus, in this case
$f_0^{\lambda}$ is represented by a function and therefore the
differential cross section is given by
\begin{equation}
\left(\frac{d\sigma}{d \theta}\right)_0^\lambda(k, \theta) = \frac{2}{\pi
k}\left|\sum_{m=-\infty}^\infty \frac{J_{|m|}(k a) - \lambda J'_{|m|}(k
a)}{ H_{|m|}^{(1)}(k a) - \lambda H_{|m|}^{(1)'}(k a)} e^{i m \theta}
\right|^2.
\end{equation}

On the other hand, again by \eqref{ampesp}, the scattering amplitude
associated with $H^\lambda$, with non-zero magnetic field, that is,
$0<\alpha<1$, is given by
\begin{equation}
f_\alpha^\lambda (k, \theta) = \frac{1}{(2\pi i k
)^{1/2}}\sum_{m=-\infty}^\infty \left( e^{2 i \delta_m^\lambda(k,\alpha)}
-1 \right)e^{i m \theta},
\end{equation}
and since
\begin{equation}
e^{2 i \delta_m^{\lambda}(k,\alpha)} = - e^{2 i \Delta_m(\alpha)}\left[
\frac{ H_{|m+\alpha|}^{(2)}(k a) - \lambda H_{|m+\alpha|}^{(2)'}(k a)} {
H_{|m+\alpha|}^{(1)}(k a) - \lambda H_{|m+\alpha|}^{(1)'}(k a)}\right],
\end{equation}
we obtain
\begin{multline}
f_\alpha^\lambda (k, \theta) = \frac{1}{(2\pi i k )^{1/2}} \\
\times \sum_{m=-\infty}^\infty \left( - e^{2 i \Delta_m(\alpha)}\left[
\frac{ H_{|m+\alpha|}^{(2)}(k a) - \lambda H_{|m+\alpha|}^{(2)'}(k a)} {
H_{|m+\alpha|}^{(1)}(k a) - \lambda H_{|m+\alpha|}^{(1)'}(k a)}\right] -1
\right)e^{i m \theta}.
\end{multline}
Now, let $n\in\Z$ be fixed and  change  variable $m'= m+n$ in the
summation index. Then $m = m'-n$ and since
$\Delta_m(\alpha)=(\pi/2)(|m|-|m+\alpha|)$, we obtain
\begin{multline}
f_\alpha^\lambda (k, \theta) = \frac{e^{-i n \theta}}{(2\pi i k )^{1/2}}
\sum_{m'=-\infty}^\infty \Biggl( - e^{2 i \delta_{m'}(\alpha-n)}(-1)^n \\
\times \left.\left[ \frac{ H_{|m'+\alpha-n|}^{(2)}(k a) - \lambda
H_{|m'+\alpha-n|}^{(2)'}(k a)} { H_{|m'+\alpha-n|}^{(1)}(k a) - \lambda
H_{|m'+\alpha-n|}^{(1)'}(k a)}\right] -1 \right)e^{i m' \theta},
\end{multline}
which can be written as
\begin{equation}
f_\alpha^\lambda (k, \theta) = (-1)^n e^{-i n \theta} f_{\alpha-n}^\lambda
(k, \theta) + (2\pi / ik )^{1/2}[(-1)^n - 1]\delta(\theta),\quad n\in\Z.
\end{equation}
Thus, the differential cross section for the Robin self-adjoint extension
of the initial AB hamiltonian is given by ($\theta\neq 0$)
\begin{multline}
\left(\frac{d\sigma}{d \theta}\right)_\alpha^\lambda(k, \theta) =
\frac{1}{2\pi k} \\
\times \left|\sum_{m=-\infty}^\infty \left( e^{2 i \Delta_m(\alpha)}\left[
\frac{ H_{|m+\alpha|}^{(2)}(k a) - \lambda H_{|m+\alpha|}^{(2)'}(k a)} {
H_{|m+\alpha|}^{(1)}(k a) - \lambda H_{|m+\alpha|}^{(1)'}(k a)}\right] + 1
\right)e^{i m \theta} \right|^2,
\end{multline}
which is periodic in $\alpha$ with period $1$. This is a justification for
the restriction $0\le\alpha<1$. It is convenient to write
\begin{equation}
f_{\alpha}^{\lambda}(k,\theta) = f_{\alpha}(k,\theta) +
f_{r,\lambda}(k,\theta),
\end{equation}
where $f_\alpha$ is the scattering amplitude of the case of radius zero
$a=0$ with Dirichlet condition at the origin, which was discussed above,
\begin{equation}
f_{\alpha}(k, \theta) = (2\pi i k )^{-1/2}\sum_{m=-\infty}^\infty \left(
e^{2 i \Delta_m(\alpha)} -1 \right)e^{i m \theta},
\end{equation}
and with $f_{r,\lambda}$ given by
\begin{equation}
f_{r,\lambda}(k,\theta) = -\left(\frac{2}{\pi i
k}\right)^{1/2}\sum_{m=-\infty}^\infty e^{2 i
\Delta_m(\alpha)}\frac{J_{|m+\alpha|}(k a) -\lambda J'_{|m+\alpha|}(k a)
}{H_{|m+\alpha|}^{(1)}(k a)  - \lambda H_{|m+\alpha|}^{(1)'}(k a)} e^{i m
\theta}.
\end{equation}
By the same argument presented above, the series for
$f_{r,\lambda}$ is convergent, and  $f_\alpha(k,\theta)$ is given by
\eqref{exprsFalpha}.

Therefore, the differential cross section for the Robin extension with
parameter $\lambda$, for
$k\neq 0$ and $\theta\neq 0$, is given by
\begin{multline}
\left( \frac{d\sigma}{d\theta} \right)_{\alpha}^{\lambda}(k,\theta) =
\left|\frac{\sin(\pi\alpha)}{(2\pi i k)^{1/2}} \frac{e^{-
i\theta/2}}{\sin(\theta/2)} \right.   \\
\left. -\left(\frac{2}{\pi i k}\right)^{1/2}\sum_{m=-\infty}^\infty e^{2 i
\Delta_m(\alpha)}\frac{J_{|m+\alpha|}(k a) -\lambda J'_{|m+\alpha|}(k a)
}{H_{|m+\alpha|}^{(1)}(k a)  - \lambda H_{|m+\alpha|}^{(1)'}(k a)} e^{i m
\theta}\right|^2.
\end{multline}
Again, $\lambda=0$ corresponds to the Dirichlet case, whereas
$\lambda=\infty$ to the Neumann boundary condition.

\subsection{Scattering comparison}
In this section we present some figures and comments to illustrate and
compare the scattering results obtained in the previous subsections. In
the figures, we have fixed the value of the solenoid radius to~$a=1$. Due
to the symmetry of the differential cross
sections as function of $\theta$, the corresponding plots are presented
only for  $0<\theta\le \pi$.
Recall that the scattering in case $\alpha=0$ is simply due to the
solenoid of non-zero radius, and one notices the Aharonov-Bohm effect by
comparing  this case with the scattering for different values of~$\alpha$
(in particular for non-integer~$\alpha$).

In the following, we collect the expressions for the scattering operators
for Dirichlet, Neumann and Robin extensions,  respectively,
\begin{multline}
S_{\alpha,m}^\mathcal D = \frac { \cos\beta \left[ N_{|m+\alpha|}(ka)^2 -
J_{|m+\alpha|}(ka)^2 \right] - 2\,\sin\beta\, J_{|m+\alpha|}(ka)
N_{|m+\alpha|}(ka) } { N_{|m+\alpha|}(ka)^2 + J_{|m+\alpha|}(ka)^2} \\
 + i\, \frac{\sin\beta \left[ N_{|m+\alpha|}(ka)^2 - J_{|m+\alpha|}(ka)^2
\right] + 2\cos\beta\, J_{|m+\alpha|}(ka) N_{|m+\alpha|}(ka)
}{N_{|m+\alpha|}(ka)^2 + J_{|m+\alpha|}(ka)^2},
\end{multline}
\begin{multline}
S_{\alpha,m}^\mathcal N = \frac { \cos\beta \left[ N_{|m+\alpha|}'(ka)^2 -
J_{|m+\alpha|}'(ka)^2 \right] - 2\,\sin\beta\, J_{|m+\alpha|}'(ka)
N_{|m+\alpha|}'(ka) } { N_{|m+\alpha|}'(ka)^2 + J_{|m+\alpha|}'(ka)^2} \\
+ i\, \frac{\sin\beta \left[ N_{|m+\alpha|}'(ka)^2 - J_{|m+\alpha|}'(ka)^2
\right] + 2\cos\beta\, J_{|m+\alpha|}'(ka) N_{|m+\alpha|}'(ka)
}{N_{|m+\alpha|}'(ka)^2 + J_{|m+\alpha|}'(ka)^2},
\end{multline}
and
\begin{align}
\begin{split}
& S_{\alpha,m}^\lambda = \frac { \cos\beta \left[\left( N_{|m+\alpha|}(ka)
- \lambda N_{|m+\alpha|}'(ka)\right)^2 - \left( J_{|m+\alpha|}(ka) -
\lambda J_{|m+\alpha|}'(ka) \right)^2 \right]}{\left( N_{|m+\alpha|}(ka) -
\lambda N_{|m+\alpha|}'(ka)\right)^2 + \left( J_{|m+\alpha|}(ka) - \lambda
J_{|m+\alpha|}'(ka) \right)^2} \\
& - \frac{ 2\,\sin\beta\, \left( J_{|m+\alpha|}(ka) - \lambda
J_{|m+\alpha|}'(ka)\right) \left( N_{|m+\alpha|}(ka) - \lambda
N_{|m+\alpha|}'(ka)\right) } {\left( N_{|m+\alpha|}(ka) - \lambda
N_{|m+\alpha|}'(ka)\right)^2 + \left( J_{|m+\alpha|}(ka) - \lambda
J_{|m+\alpha|}'(ka) \right)^2} \\
& + i\, \left[ \frac { \sin\beta \left[\left( N_{|m+\alpha|}(ka) - \lambda
N_{|m+\alpha|}'(ka)\right)^2 - \left( J_{|m+\alpha|}(ka) - \lambda
J_{|m+\alpha|}'(ka) \right)^2 \right]}{\left( N_{|m+\alpha|}(ka) - \lambda
N_{|m+\alpha|}'(ka)\right)^2 + \left( J_{|m+\alpha|}(ka) - \lambda
J_{|m+\alpha|}'(ka) \right)^2}\right. \\
& \left. + \frac{ 2\cos\beta\, \left( J_{|m+\alpha|}(ka) - \lambda
J_{|m+\alpha|}'(ka)\right) \left( N_{|m+\alpha|}(ka) - \lambda
N_{|m+\alpha|}'(ka)\right) } {\left( N_{|m+\alpha|}(ka) - \lambda
N_{|m+\alpha|}'(ka)\right)^2 + \left( J_{|m+\alpha|}(ka) - \lambda
J_{|m+\alpha|}'(ka) \right)^2}\right],
\end{split}
\end{align}
and recall that $\beta=\pi(|m|-|m+\alpha|)$.

\begin{figure}[p]
\hspace{-1.2cm}
\includegraphics[scale=0.60]{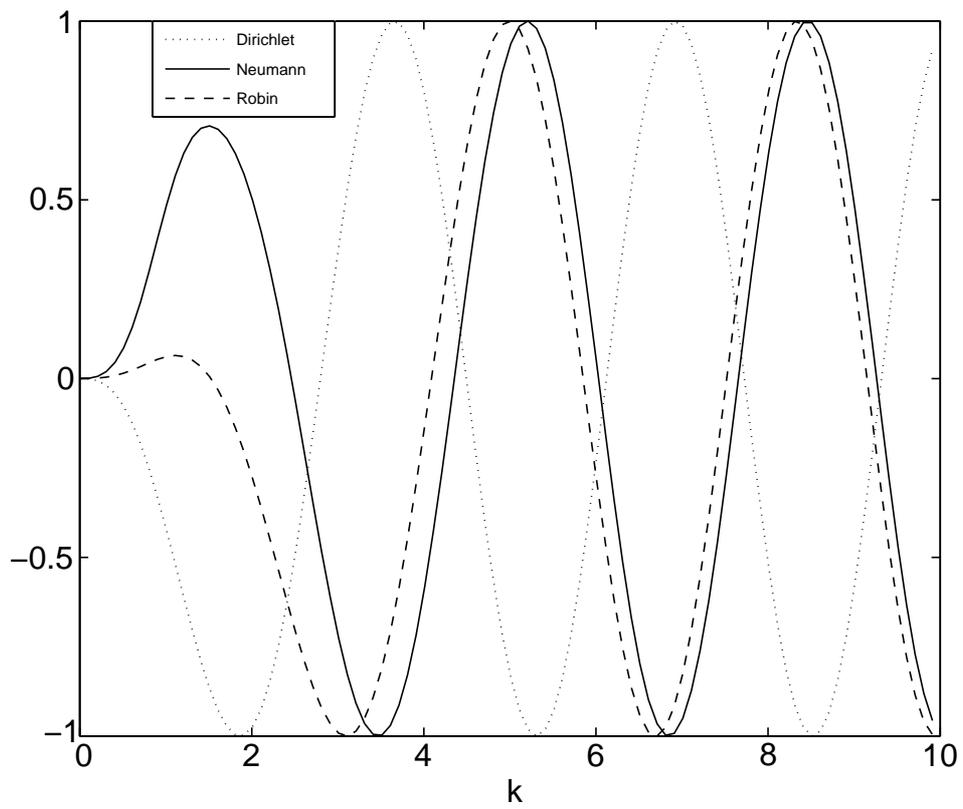}
\caption{Real parts of the scattering operators of the three extensions
with non-zero field ($\alpha = 1/2$) as function of $k$ with $a = 1$, $m =
1$ and $\lambda = 1$.}
\label{figure1}
\end{figure}

Figure~\ref{figure1} presents the real parts of
scattering operators for
the above three extensions, in a case with $\alpha\ne0$. Note that, for
high energies, the curve of the Robin scattering operator approaches the
curve of the Neumann case, and  it is evident the phase difference between
the Dirichlet and Neumann cases; this agrees with the theoretical results
of Subsection~\ref{compassint}. For very low energies ($ka\to0$), this
figure illustrates
what we have said in the last paragraph of Subsection \ref{compassint}
about the behaviour of the scattering
operator (since  $\cos\beta=0$), that is, for low energies the scattering
operator is very similar in all cases we have considered. Similar results
hold  when  no magnetic field is present, i.e., $\alpha=0$.

For each of such self-adjoint extensions, we have numerically checked
(with plots) that the scattering operators, with non-zero magnetic fields
(i.e., for any $\alpha\ne0$), approach the corresponding scattering
operators with no magnetic field (i.e., $\alpha=0$) for high energies (not
shown). Hence, given one of those self-adjoint extensions, for high
energies the scattering operator is not able to discern the presence of
magnetic field inside the solenoid or not. We note that such behaviours of
the scattering operators were found to be independent of the values of
$m$, $0<\alpha<1$, and $\lambda> 0$.

Now, we consider the important concept of differential cross section, in
the case of cylindrical solenoids of positive radius $a> 0$, and for the
 Dirichlet, Neumann and Robin extensions. The respective  expressions we
have obtained  are, for $\theta\ne0$,
\begin{multline}
\left( \frac{d\sigma}{d\theta} \right)_{\alpha}^{\mathcal D}(k,\theta)=
\left| \frac{\sin(\pi\alpha)}{(2\pi i k)^{1/2}} \frac{e^{-
i\theta/2}}{\sin(\theta/2)} \right. \\
\left.- \left(\frac{2}{\pi i k}\right)^{1/2}\sum_{m=-\infty}^\infty e^{2 i
\Delta_m(\alpha)}\frac{J_{|m+\alpha|}(k a) }{H_{|m+\alpha|}^{(1)}(k a)}
e^{i m \theta} \right|^2,
\end{multline}
\begin{multline}
\left( \frac{d\sigma}{d\theta} \right)_{\alpha}^{\mathcal N}(k,\theta)=
\left| \frac{\sin(\pi\alpha)}{(2\pi i k)^{1/2}} \frac{e^{-
i\theta/2}}{\sin(\theta/2)}\right. \\
\left. -\left(\frac{2}{\pi i k}\right)^{1/2}\sum_{m=-\infty}^\infty e^{2 i
\Delta_m(\alpha)}\frac{J'_{|m+\alpha|}(k a) }{H_{|m+\alpha|}^{(1)'}(k a)}
e^{i m \theta} \right|^2,
\end{multline}
and
\begin{multline}
\left( \frac{d\sigma}{d\theta} \right)_{\alpha}^{\lambda}(k,\theta) =
\left| \frac{\sin(\pi\alpha)}{(2\pi i k)^{1/2}} \frac{e^{-
i\theta/2}}{\sin(\theta/2)} \right. \\
\left. - \left(\frac{2}{\pi i k}\right)^{1/2}\sum_{m=-\infty}^\infty e^{2
i \Delta_m(\alpha)}\frac{J_{|m+\alpha|}(k a) -\lambda J'_{|m+\alpha|}(k a)
}{H_{|m+\alpha|}^{(1)}(k a)  - \lambda H_{|m+\alpha|}^{(1)'}(k a)} e^{i m
\theta} \right|^2.
\end{multline}

\begin{figure}[p]
\hspace{-1.2cm}
\includegraphics[scale=0.60]{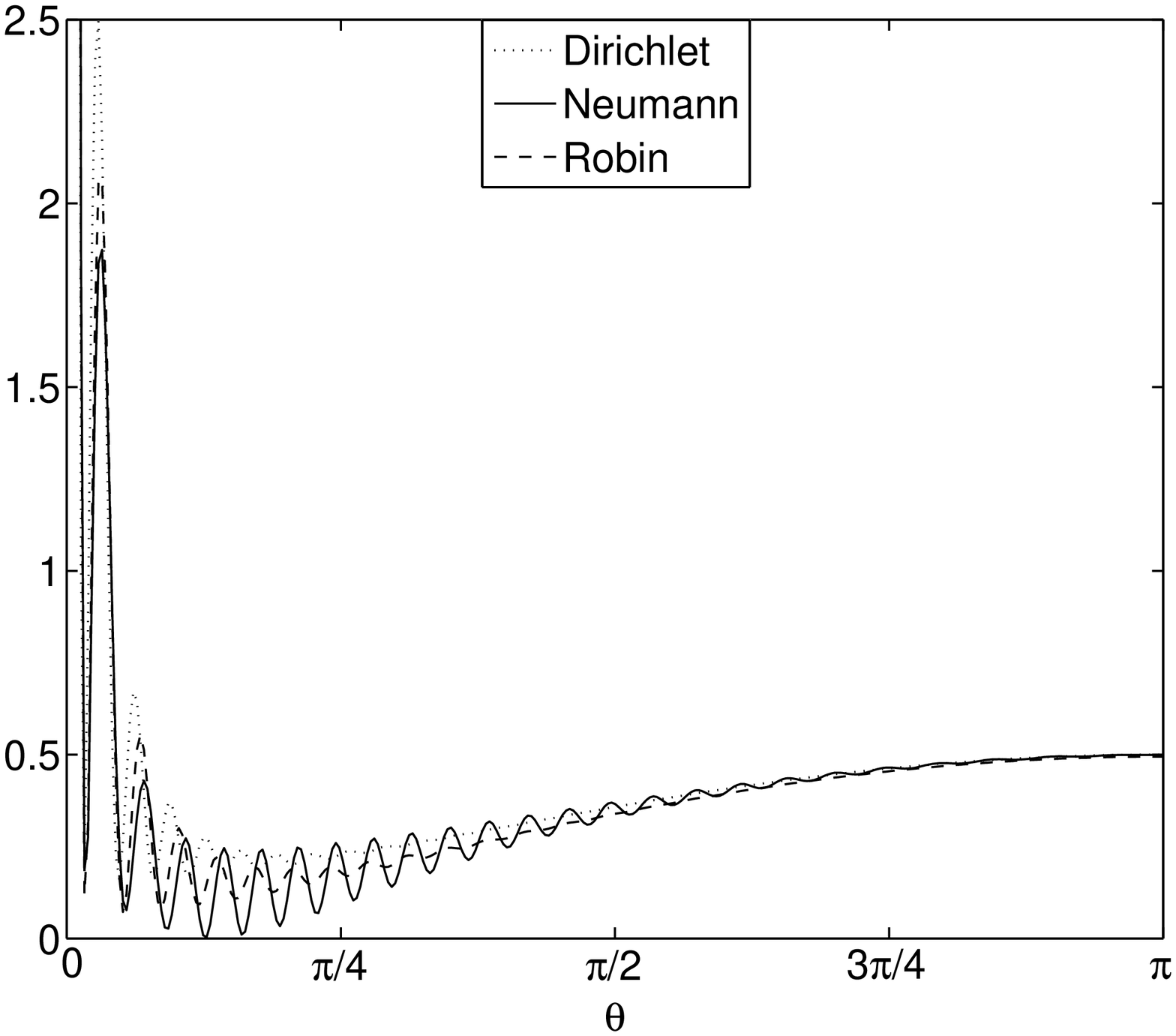}
\caption{Differential cross section as function of $\theta$ in the case
with field ($\alpha=1/2$), with $a=1$, $k=30$ and $\lambda=1/10$.}
\label{figure2}
\end{figure}

For high energies, we have found that the differential cross section of
Neumann and Robin cases are very close to Dirichlet for each given
$0\le\alpha<1$, except in a neighborhood of $\theta = 0$ and $\theta =
2\pi$. Figure~\ref{figure2} shows  those curves for $\alpha =
1/2$.

\begin{figure}[p]
\hspace{-1.2cm}
\includegraphics[scale=0.60]{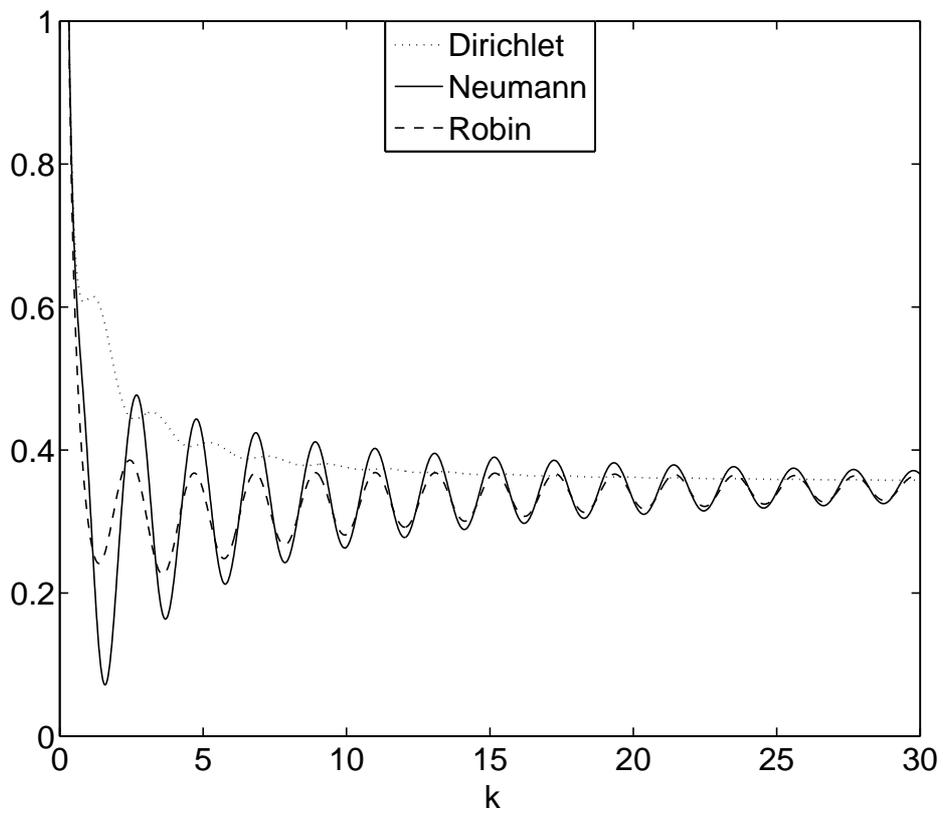}
\caption{Differential cross section as function of $k$ in the case with
field ($\alpha=1/2$), with $a=1$, $\theta=\pi/2$ and $\lambda=1$.}
\label{figure3}
\end{figure}

Figure~\ref{figure3} shows the differential cross section of the
three extensions in
terms of the ``energy'' $k$, for the case with non-zero field, represented
by $\alpha=1/2$, fixed angle $\theta = \pi/2$ and $\lambda = 1$; note the
different behaviours for high and low energies.

\begin{figure}[p]
\hspace{-1.2cm}
\includegraphics[scale=0.60]{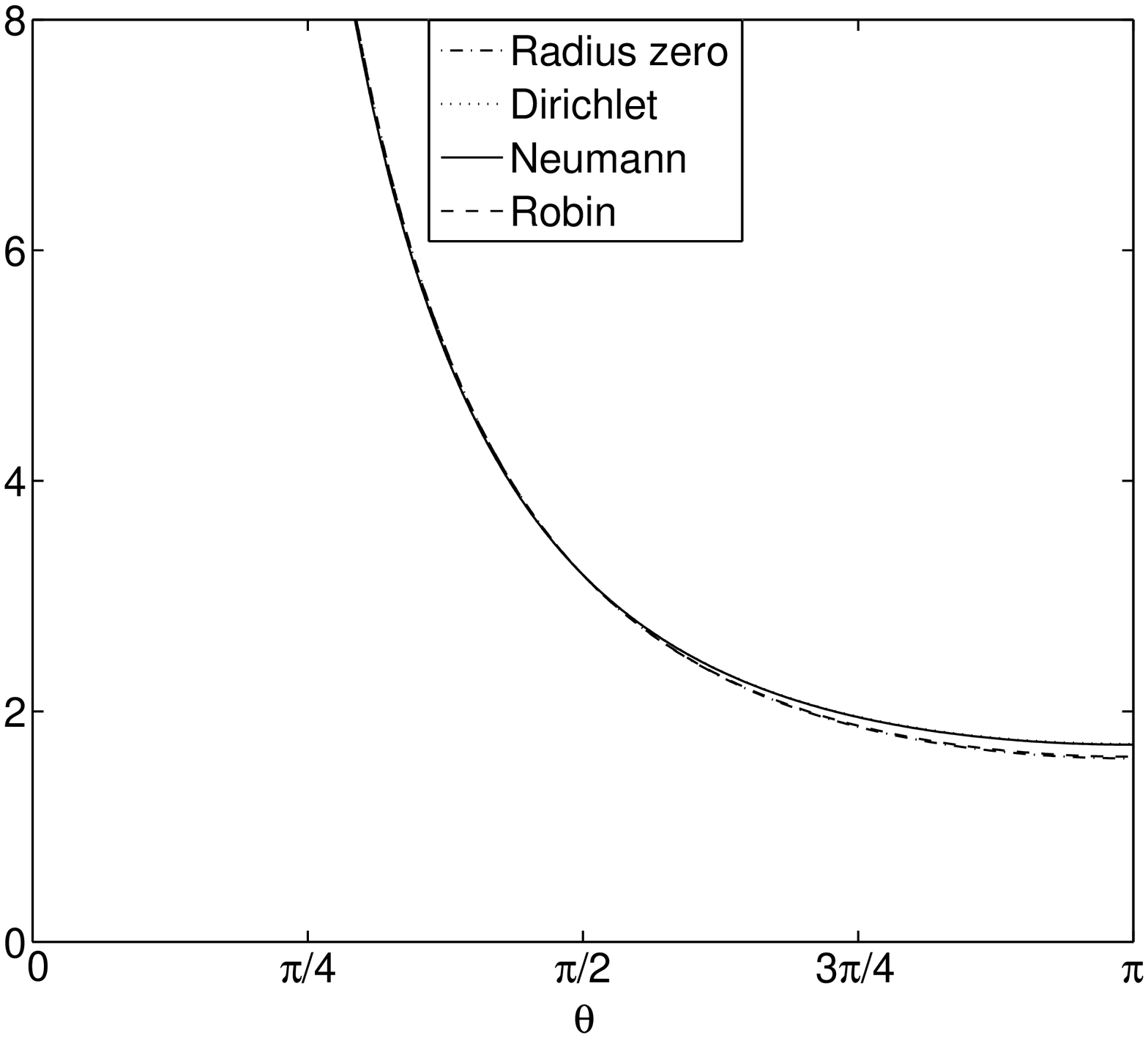}
\caption{Differential cross section as function of $\theta$ in the case
with field ($\alpha=1/2$), with $a=1$, $k=1/10$ and $\lambda=1$.}
\label{figure4}
\end{figure}

For $k\to 0$, in the case with field ($\alpha\neq 0$) and positive radius,
we have found that the differential cross sections for the three cases
have the same behaviour, which is approximately given by the differential
cross section of the case with zero radius \eqref{scr0} and Dirichlet
condition at the origin. See
Figure~\ref{figure4}.

\begin{figure}[p]
\hspace{-1.2cm}
\includegraphics[scale=0.60]{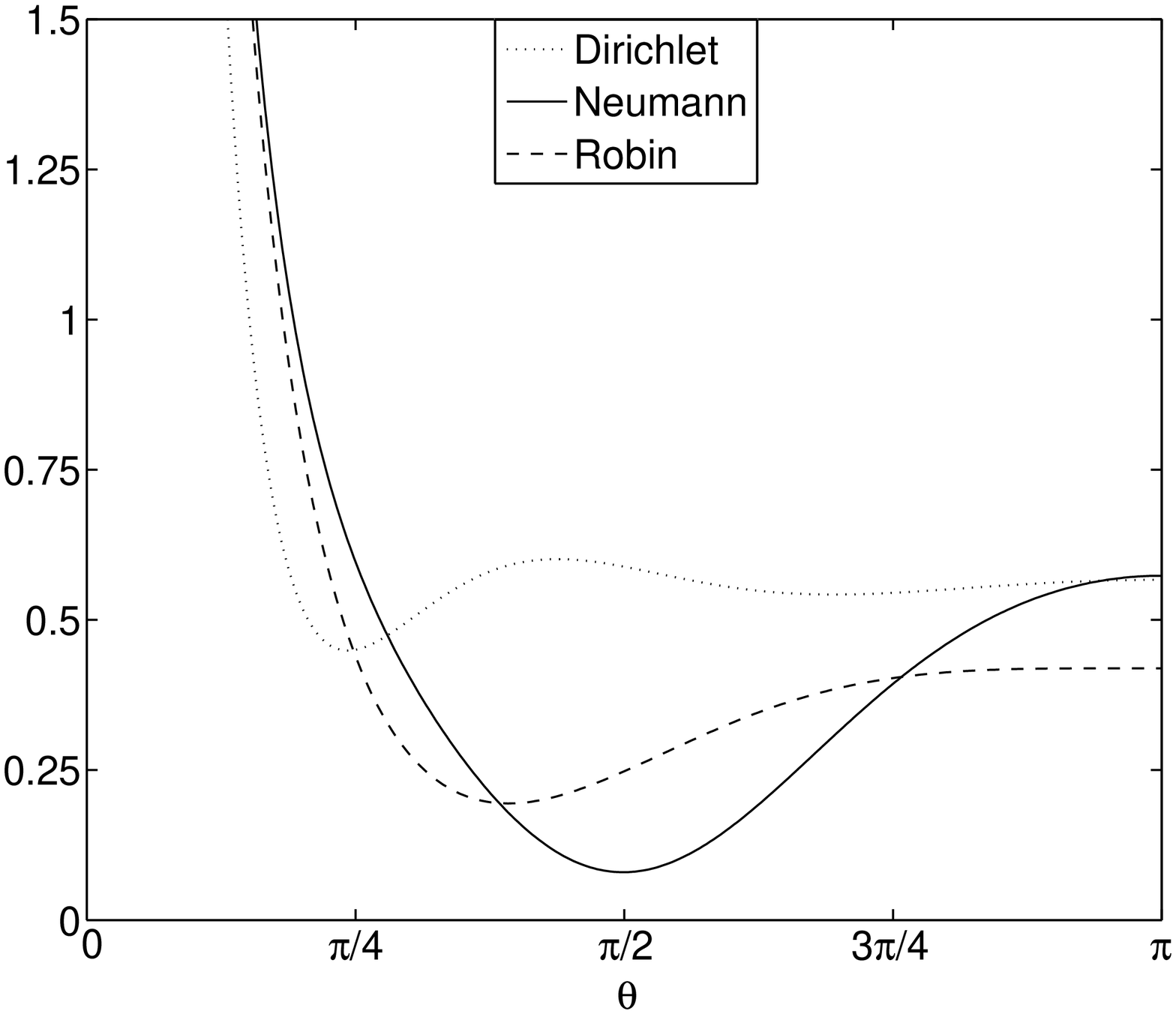}
\caption{Differential cross section as function of $\theta$ in the case
with field ($\alpha=1/2$), with $a=1$, $k=3/2$ and $\lambda=1$.}
\label{figure5}
\end{figure}

For intermediate energies the differential cross sections of the
extensions differ significantly, as illustrated in
Figure~\ref{figure5}; this seems interesting, since it is an explicitly
distinction among different boundary conditions.

Finally, we mention that for small  $\lambda$ (for example, $\lambda =
1/10$), the differential cross section for the Robin extension approaches
the values obtained for the Dirichlet case, and when we choose $\lambda$
large (for example, $\lambda = 10$)  the values for the Neumann extension
are virtually recovered. This is certainly expected.

\section{Conclusions}
With respect to the mathematical problems related to the traditional
magnetic AB setting, that is, the one associated with an infinitely long
solenoid, in this work we have based our investigations on two
cornerstones. First, we have considered  the more realistic case of a
solenoid of positive radius $a>0$; and second, we did not take for granted
that the boundary conditions on the solenoid border $\mathcal S$ is
Dirichlet (although there are physical insight \cite{Ruij} and
mathematical arguments that support this choice \cite{deOP}).

The boundary conditions that are physically compatible with quantum
mechanics are those that define self-adjoint extensions of the initial AB
hamiltonian \eqref{initialABh}. We have characterized all
such self-adjoint extensions whose domains are contained in the natural
Sobolev space $\hil^2(\mathcal S')$; this was done via boundary triples,
and our main contribution was the inclusion of the vector
potential in the operator action, by taking into account the symmetry of
the problem, and a gauge choice as well, to simplify expressions.

The important cases of Dirichlet, Neumann and Robin are among the
self-adjoint extensions we have characterized via boundary triples, and
the next step was to
study the scattering for these self-adjoint hamiltonians; such study was
based on \cite{Ruij}, where the particular case of Dirichlet boundary
condition was considered. For some parameter ranges, that is,
$0\le\lambda\le\infty$, we have proven that  the
wave operators are well defined and complete; furthermore the hamiltonian
is positive and has no eigenvalues. We remark that for negative values of
$\lambda$ one can not discard the presence of eigenvalues (see, for
instance, Exercise~7.3.3 in~\cite{ISTQD}), and so bounded states could
emerge from the Robin boundary condition; this is an interesting
possibility we think it is worth investigating.

Then we have explicitly
calculated the scattering operators and subsequent scattering cross
sections, and they were our natural physical quantities used to compare
different self-adjoint extensions. Note that the scattering cross section
is a distribution in general, but for the scattering angle $\theta\ne0$ it
is represented by a continuous function ($k\ne0$).

For high energies, we have found that the scattering operator for the
Robin case is similar to the Neumann one, but different from the Dirichlet
case. On the other hand, for low energies the behaviour of the scattering
operator is independent of these self-adjoint extensions. Such results hold
for each fixed $0\le\alpha<1$.

With respect to the differential cross section, for ``intermediate
energies'' its behaviour depends significantly on the choice among the
three self-adjoint
extensions we have considered.

To finish, we underline that, in general, our scattering results depend on
the magnetic field parameter
$\alpha$, and this is actually a confirmation of the presence of the AB
effect (when $0<\alpha<1$) in different self-adjoint extensions!

\subsubsection*{Acknowledgments} {\small CRdeO thanks partial
support from CNPq (Brazil), and MP acknowledges partial support from CNPq
and Funda\c c\~ao Arauc\'aria (Brazil).}

\end{document}